\documentclass{ectj}
\usepackage{amsfonts,amssymb,graphics,epsfig,verbatim,bm,latexsym,amsmath,url,amsbsy}
\usepackage{adjustbox}
\usepackage{xurl}
\usepackage{mathtools}
\usepackage{float}
\usepackage{natbib}

\usepackage{threeparttable}
\usepackage{lscape}
\usepackage{scrextend}
\usepackage{relsize}
\usepackage{multicol}
\usepackage{chngcntr}
\usepackage{etoolbox}
\usepackage{xr}

\usepackage{tabularx}

\usepackage{multirow}
\usepackage{booktabs}
\usepackage{array}
\usepackage{color}
\usepackage{afterpage}
\usepackage{framed}
\usepackage{float}

\usepackage{scalerel,stackengine}
\usepackage{autobreak}

\newtheorem{theorem}{Theorem}
\newtheorem{assumption}{Assumption}

\newtheorem{corollary}{Corollary}

\newtheorem{remark}{Remark}
\newtheorem{algorithm}{Algorithm}

\renewcommand{\thesection}{\arabic{section}}
\renewcommand{\theequation}{\arabic{section}.\arabic{equation}}

\newcounter{bean}

\received{November 2021}
\accepted{...}
\volume{20}

\makeatletter
\newcommand*{\addFileDependency}[1]{
  \typeout{(#1)}
  \@addtofilelist{#1}
  \IfFileExists{#1}{}{\typeout{No file #1.}}
}
\makeatother

\newcommand*{\myexternaldocument}[1]{%
    \externaldocument{#1}%
    \addFileDependency{#1.tex}%
    \addFileDependency{#1.aux}%
}

\myexternaldocument{Appendix}

\title[Distribution Regression in Duration Analysis]{Distribution Regression in Duration Analysis: \\
an Application to Unemployment Spells}

\author[Delgado, Garc\'{\i}a-Suaza, and Sant'Anna]{Miguel~A.~Delgado$^{\dagger}$,
                         Andr\'{e}s~Garc\'{\i}a-Suaza$^{\ddagger}$ and Pedro~H.C.~Sant'Anna$^{\dagger\dagger}$}

\address{$^{\dagger}$Universidad Carlos III de Madrid, Calle Madrid 126, 28903 Getafe, Madrid, Spain}
\email{delgado@est-econ.uc3m.es}

\address{$^{\ddagger}$Universidad del Rosario, Calle 14 625, Bogot\'a, Colombia.}
\email{andres.garcia@urosario.edu.co}

\address{$^{\ddagger}$ Microsoft and Vanderbilt University, 415 Calhoun Hall, Nashville, TN 37240, USA.}
\email{pedro.h.santanna@vanderbilt.edu}

\def\AmSTeX{$\cal A$\kern-.1667em\lower.5ex\hbox{$\cal M$}\kern-.125em
            $\cal S$-\TeX}
\def\BibTeX{{\rm B\kern-.05em{\sc i\kern-.025em b}\kern-.08em
            T\kern-.1667em\lower.7ex\hbox{E}\kern-.125emX}}

\begin{document}

\begin{abstract}
This article proposes inference procedures for distribution regression
models in duration analysis using randomly right-censored data. This
generalizes classical duration models by allowing situations where
explanatory variables' marginal effects freely vary with duration time. The
article discusses applications to testing uniform restrictions on the
varying coefficients, inferences on average marginal effects, and others
involving conditional distribution estimates. Finite sample properties of
the proposed method are studied by means of Monte Carlo experiments.
Finally, we apply our proposal to study the effects of unemployment benefits
on unemployment duration.


\end{abstract}

\section{Introduction}
\setcounter{equation}{0}  

Existing semiparametric duration models can be broadly classified into two
groups: those based on the conditional hazard and those based on the
quantile regression. The former includes the proportional hazard model (\citealp{Cox1972, Cox1975}), the proportional odds model (\citealp{Clayton1976, Bennett1983, Murphy1997}), and the accelerated failure time model (\citealp{Kalbfleisch1980}); see \cite{Guo2014} for an overview. In these models, the conditional
hazard identifies the conditional cumulative distribution expressed in terms
of the error's marginal distribution of a transformed failure time
regression model (see, e.g., \citealp{Hothorn2014}). Censored quantile regression proposals are relatively more recent, see, e.g., 
\cite{Ying1995}, \cite{Honore2002}, \cite{Portnoy2003}, \cite{Peng2008}, and 
\cite{Wang2009c}.

Conditional hazard and quantile regression models are alternative modeling
strategies with advantages and drawbacks. Classical conditional hazard
specifications impose identification conditions that are difficult to justify
in some circumstances. For instance, the proportional hazard specification
rules out important forms of heterogeneity, see, e.g., \cite{Portnoy2003}
and the discussion in Section \ref{motivation}. On the other hand, models
based on quantile regression specifications avoid this problem, but impose
that the underlying conditional cumulative distribution of duration time is
absolutely continuous, which rules out other sampling schemes such as discrete outcomes.

The distribution regression approach was proposed by \cite{Foresi1995} and
formalized by \cite{Chernozhukov2013a}; see also \cite{Rothe2013a, Rothe2016a},  \cite{Chernozhukov2019} and \cite{Chernozhukov2018a} for later
developments. One attractive feature of this modeling strategy is that marginal effects associated with distribution
regression models are more flexible than those obtained from classical
conditional hazard specifications, as they can accommodate richer forms of
heterogeneity. In contrast to quantile regression, distribution regression
can accommodate discrete, continuous and mixed duration data in a unified
manner under fairly weak regularity conditions. \cite{Chernozhukov2013a} show that distribution regression encompasses the \cite{Cox1972} model as a special case and represents a useful alternative to quantile regression.

The distribution regression model specifies the distribution of the duration outcome variable by means of a generalized linear model with known link function and nonparametric varying coefficient depending on duration time. When data is uncensored, the varying coefficients at each duration value are consistently estimated by binary regressions. This method is also valid using censored data with known censoring time, which is an unlikely situation in duration studies, e.g., unemployment duration spells. When duration and censoring times are potentially unobserved, the standard binary regression approach lead to inconsistent estimators.

In this article, we propose a weighted binary regression procedure using  \cite{Kaplan1958} weights. The varying coefficients are identified by means of a set of moment restrictions of the duration time and covariates' joint distribution. These moments are consistently estimated by their corresponding Kaplan-Meier integrals when the duration time is subjected to censoring; if censoring is not an issue, the Kaplan-Meier integrals reduce to the sample analogue of the moment restrictions. The resulting estimator admits a representation as a non-degenerate U-statistic of order three, whose H\'{a}jek projection is asymptotically distributed as a normal with an asymptotic variance that can be consistently estimated from the data. The resulting inference procedures on the coefficients do not rely on choosing tuning parameters such as bandwidths, imposing smoothness conditions on the censoring random variable, or using truncation arguments.

We establish the consistency and finite-dimensional distribution convergence
of the varying coefficients under weak regularity conditions. Under more
restrictive assumptions we justify the convergence in distribution of the
estimated coefficient function as a random element of a suitable metric
space. We also justify bootstrap-based inference procedures.

In order to justify the asymptotic inferences based on our proposed method, we impose restrictions on the joint distribution of the duration time, censoring time and covariates. \citet{Stute1993, Stute1996a} provided strong consistency and a central limit theorem for Kaplan-Meier integrals that estimate moments of (partially observed) durations time and covariates under the following identification conditions: (a) the duration and censoring times are independent, and (b) the event of being censored is conditionally independent of the covariates given the duration time. Stute shows that the two conditions suffice to identify the joint distribution of duration time and covariates from the observed censored data, and, hence, to identify corresponding moments of these conditional distributions. Of course, the identification of alternative models may require alternative regularity conditions. For instance, identification of the \citet{Cox1972}'s hazard model, \citet{Aalen1980}'s additive hazard model, or censored quantile regressions models (\citealp{Portnoy2003}) only requires that duration and censoring times are independent given covariates. However, as we mention in Section 2, these models usually exclude important types of heterogeneity, or restrict their attention to absolutely continuous duration data. Alternatively, one can potentially estimate the moment equations using inverse probability weighting (IPW) methods, as suggested by \citet{Robins1992, Robins1995a} and later extended by \citet{Wooldridge2007a} to general missing data problems. Although this path allows one to only assume that duration and censoring times are independent given covariates, it also requires that the conditional distribution of the censoring times given covariates is (uniformly) bounded away from one, which rules out discrete distributions and the most popular distribution functions used in survival analysis such as exponential, Gamma, log-normal and Weibull, among others. If one is not comfortable with these additional restrictions, truncation/trimming arguments need to be carefully introduced, and the choice of tuning parameters becomes a first-order concern. By adopting the Kaplan-Meier integrals approach, we avoid these alternative restrictions, but rely on conditions (a) and (b) above.

We illustrate the relevance of our proposal by assessing how changes in
unemployment insurance benefits affect, on average, the distribution of
unemployment duration, using data from the Survey of Income and Program
Participation for the period 1985-2000. We find that, by allowing the
distributional marginal effects to vary over the unemployment spell, our
proposed method can reveal interesting insights when compared to traditional
hazard models as those used by \cite{Chetty2008}. For instance, our results
suggest a non-monotone marginal effect of an increase in unemployment
insurance on the unemployment duration distribution. Such a finding is in
sharp contrast with those obtained using classical proportional hazard
models. On the other hand, our results agree with \cite{Chetty2008} in that
increases in unemployment insurance have larger effects on
liquidity-constrained workers, suggesting that unemployment insurance
affects unemployment duration not only through a moral hazard channel but
also through a liquidity effect channel.

The rest of the article is organized as follows. Section \ref{motivation}
introduces the basic notation and motivates distribution regression models using duration data as an alternative to classical conditional hazard modeling. Section \ref{DRTrad} describes our
estimation procedure, whereas Section \ref{Asy} introduces regularity conditions needed to justify inferences on the
distribution regression varying coefficients. All the results are proved in
the Supplementary Appendix. Application of the results to
different contexts are placed in Section \ref{theory_application}. Section \ref{simulation} briefly summarizes the results of the Monte Carlo simulations -- detailed discussion is presented in the Supplementary Appendix\footnote{The Supplementary Appendix is available at \url{https://pedrohcgs.github.io/files/Delgado_GarciaSuaza_SantAnna_2021_EJ-Appendix.pdf}}. Finally,
we apply the proposed techniques to investigate the effect of unemployment
benefits on unemployment duration in Section \ref{Application}.   All our replication files are available at \url{https://github.com/pedrohcgs/KMDR-replication}.

\section{Distribution Regression with Duration Outcomes} \label{motivation}
\setcounter{equation}{0}

Consider the $\mathbb{R}^{1+k}-valued$ random vector $\left( T,X\right) $
defined on $\left( \Omega ,\mathcal{F},\mathbb{P}\right) ,$ where $T$ is the
duration outcome of interest and $X$ is a $k-$dimensional vector of
time-invariant covariates, with supports $\mathcal{T}$ and $\mathcal{X},$
respectively. Henceforth, let $\mathbf{X} = (1,X')'$ and $\mathbf{x} = (1,x')'$. We assume that the conditional cumulative distribution
function (CDF) of $T$ given $X$ follows the distribution regression (DR)
model, 
\begin{equation}
	F_{\left. Y\right\vert X}\left( \left. t\right\vert X\right) =\Lambda \left(
	\varphi \left( t\right) +X^{\prime }\beta _{0}\left( t\right) \right) \text{
		for some }\left( \varphi \left( t\right) ,\beta _{0}\left( t\right) \right)
	\in \Theta \text{ a.s.}  \label{dr}
\end{equation}%
where $\theta _{0}\left( \cdot \right) =\left( \varphi \left( \cdot \right)
,\beta _{0}\left( \cdot \right) ^{\prime }\right) ^{\prime }\mapsto \Theta
\subseteq \mathbb{R}^{k+1}$ is a vector of nonparametric functions, and $\Lambda$ is a known link function.

\cite{Chernozhukov2013a} point out that the DR model is a flexible alternative to classical duration models as it allows coefficients to vary with duration time. In particular, they show that it nests the \cite{Cox1972} proportional hazard (PH) model. Model (\ref{dr}) also generalizes other traditional models commonly adopted in
duration analysis such as \cite{Kalbfleisch1980} accelerated failure time (AFT) model, and \cite{Clayton1976} proportional odds (PO) model.

We note that all the aforementioned classical duration models are special
cases of the linear transformation model 
\begin{equation}
	\varphi \left( T\right) =-X^{\prime }\beta _{0}+\Lambda ^{-1}(U)\text{ }a.s.,
	\label{2}
\end{equation}%
where $\beta _{0}\in \mathbb{R}^{k}$ is a vector of unknown parameters, $%
\varphi $ is a (potentially unknown) monotonically increasing transformation
function, $\Lambda ^{-1}$ is the $\Lambda ^{\prime }s$ quantile function,
and $U$ is uniformly distributed in $\left[ 0,1\right] $ independently of $X$. For instance, $(a)$ the PH model corresponds to model (\ref{2}) with $\Lambda $ the complementary log-log (cloglog) link function, $\Lambda \left(u\right) =1-\exp \left( -\exp \left( u\right) \right) $; $(b)$ the AFT model
corresponds to (\ref{2}) with $\Lambda $ commonly assumed to be the link
function of a log-logistic or of a Weibull distribution; and $(c)$ the PO
model corresponds to (\ref{2}) with $\Lambda $ the logistic link function, $%
\Lambda \left( u\right) =\exp \left( u\right) \left/ \left( 1+\exp \left(
u\right) \right) \right. $; see, e.g. \cite{Doksum1990}, \cite{Cheng1995}
and \cite{Hothorn2014}. Indeed, the CDF associated with (\ref{2}) is  given by
\begin{equation}
	F_{T|X}\left( t|X\right) =\Lambda \left( \varphi \left( t\right) +X^{\prime
	}\beta _{0}\right) .  \label{3}
\end{equation}%
Therefore, (\ref{dr}) is a natural generalization of (\ref{3}) that allows
all the slope coefficients varying with $t$. Other duration models with
varying slope coefficients are also special cases of (\ref{dr}). For
instance, \cite{Aalen1980} semi-parametric additive hazard model reduces to (%
\ref{dr}) with $\Lambda \left( u\right) =1-\exp \left( -u\right) $. The
non-proportional odds model considered by \cite{McCullagh1980} and \cite%
{Armstrong1989} can also be expressed in terms of the DR specification (\ref%
{dr}) when $\Lambda $ is the logistic link function.

We conclude this section by emphasizing that allowing for varying slope coefficients is particularly important to capture richer heterogeneous effects
of covariates across the distribution of the duration outcome. Indeed, traditional duration specifications, such as the PH, PO or AFT models,
implicitly impose that all covariates affect the CDF of $T$ in a
proportional, and monotonic form. If $\Lambda $ is differentiable with Lebesgue
density $\lambda$, partial effects for these models have the form
\begin{equation*}
	\frac{d}{dx}F_{T|X}\left( t |x\right) =\beta _{0}\lambda \left( \varphi
	\left( t \right) +x^{\prime }\beta _{0}\right).
\end{equation*}%
Thus, the sign of the marginal effect of any $X$'s component is not allowed to
vary with $t$, which may be too restrictive in some applications. For
instance, in the context of the empirical application in Section \ref{Application}, (\ref{3}) does not allow a non-monotonic effect, e.g. U-shaped, ruling out
non-stationary search models in the spirit of \cite{VandenBerg1990}. The DR
model (\ref{dr}) bypass such limitations since
\begin{equation*}
	\frac{d}{dx}F_{T|X}\left( t |x\right) =\beta _{0}(t )\lambda \left(
	\varphi \left( t \right) +x^{\prime }\beta _{0}\left( t \right)
	\right) ,
\end{equation*}%
which sign is allowed to vary with $t$. We view this added  flexibility as an attractive feature of the DR modeling approach.

\section{The Kaplan-Meier Distribution Regression}\label{DRTrad}
\setcounter{equation}{0}

The main challenge when dealing with duration data is that the outcome of
interest $T$ is subject to censoring according to a variable $C$. As so,  inferences must be
based on observations $\left\{ Y_{i},X_{i},\delta _{i}\right\} _{i=1}^{n}$,
with $Y_{i}=\min \left( T_{i},C_{i}\right) $ the realized (potentially censored) outcome, $%
\delta _{i}=1_{\left\{ T_{i}\leq C_{i}\right\} }$ indicates whether or not $T$ is observed, and $\left\{ T_{i},C_{i},X_{i}\right\} _{i=1}^{n}$ are independent and identically distributed ($iid$)
as $\left( T,C,X\right)$. Henceforth, $1_{\left\{ A\right\} }$ is the
indicator function of the event $A$.

Suppose for the moment that $\delta _{i}=1,$ i.e., $T_{i}=Y_{i},$ for all $i=1,\dots ,n$. In this case, the joint $\left( T,X'\right)'$ CDF, $F\left( t,x\right)$, is consistently estimated by its sample version $
\tilde{F}_{n}\left( t,x\right) =n^{-1}\sum_{i=1}^{n}1_{\left\{ Y_{i}\leq y,X_{i}\leq x \right\} }$, where inequalities are coordinate-wise. Therefore, following \cite{Foresi1995} and \cite{Chernozhukov2013a},
$\theta _{0}\left( t\right) $ can be consistently estimated by $\tilde{\theta}_{n}(t)$, the maximizer of the conditional likelihood function of $\left\{
1_{\left\{ Y_{i}\leq t\right\} },X_{i}\right\} _{i=1}^{n}$ 
\begin{equation}
	\tilde{Q}_{n}(\theta ,t)=\int_{\mathbb{R}^{1+k}}\ln p_{\theta ,t}(1_{\left\{
		y\leq t\right\} },x)\tilde{F}_{n}(dy,dx)=\frac{1}{n}\sum_{i=1}^{n}\ln
	p_{\theta ,t}(1_{\left\{ Y_{i}\leq t\right\} },X_{i}),  \label{11}
\end{equation}%
with 
\begin{equation*}
	p_{\theta ,t}(d,x)=\Lambda \left( \boldsymbol{x}^{\prime }\theta \right)
	^{d}\left( 1-\Lambda \left( \boldsymbol{x}^{\prime }\theta \right) \right)
	^{1-d}.
\end{equation*}%
However, when $T$ is subject to  right-censoring, i.e., $T\neq Y$,  $\tilde{F}_{n}$ is no longer a consistent estimator of $F$ and, hence, neither is $\tilde{\theta}_{n}\left( t\right) $.

Since $T$ is not always observed, $F$ and $\theta (t)$ must be identified
from the joint distribution of the observed  data on $\left( Y,X,\delta \right)$. However, this is not always possible without additional information. Henceforth,
for any random variable $\xi $, which can be $T,~C$ or $Y$, define $F_{\xi
}(t)\equiv \mathbb{P}\left( \xi \leq t\right) $ and $\tau _{\xi }\equiv\inf
\left( t:\mathbb{P}\left( \xi \leq t\right) =1\right) $, and let $A$ denote the
(possibly empty) set of $F_{T}$ jumps. Also, for any generic function $g$, $%
g(t-)\equiv\lim_{s\uparrow t}g(s)$. If no information about $T$ beyond $\tau _{Y}$
is available from the data, identification of 
\begin{equation*}
	F_{\ast }(t,x)\equiv \left\{ 
	\begin{array}{l}
		F(t,x)\text{ for }t<\tau _{Y} \\ 
		F(\tau _{Y}-,x)+1_{\left\{ \tau _{Y}\in A\right\} }\left[ F(\tau
		_{Y},x)-F(\tau _{Y}-,x)\right] \text{ for }t\geq \tau _{Y}.%
	\end{array}%
	\right.
\end{equation*}%
is the best that one can hope for; see, e.g., \cite{Tsiatis1975, Tsiatis1981}, and \cite{Stute1993}. In order to identify $F_{\ast }$ we impose the following assumptions.

\begin{assumption}
	\label{A1} $T$ and $C$ are independent.
\end{assumption}

\begin{assumption}
	\label{A2}$\delta $ and $X$ are conditionally independent given $T$.
\end{assumption}

Assumption \ref{A1} is introduced to identify $F_{\ast }\left( \cdot ,\infty
,...,\infty \right) $ (see \citealp{Tsiatis1975, Tsiatis1981} for discussion), whereas Assumption \ref{A2} is introduced by \cite{Stute1993} and allows us to incorporate covariates into the analysis. Taken together, these two assumptions imply that covariates should have no effect on the probability of being
censored once $T$ is known. Of course, Assumptions \ref{A1} and \ref{A2} hold if $C$ is
independent of $\left( T,X\right)$, but can also hold under more general circumstances. See \citet{Stute1993, Stute1996a, Stute1999} for a discussion on Assumptions \ref{A1} and  and \ref{A2}, and the identification of $F_*$ and its moments. Identification can be achieved under alternative conditions, in the context of alternative conditional survival models, or alternative estimation procedures. See Introduction for further comments.

\begin{remark}
	\label{Rem1}\textnormal{Under Assumptions \ref{A1} and \ref{A2}, $\tau _{Y}=\min \left(
		\tau _{T},\tau _{C}\right) $. Therefore, $F=F_{\ast }$ if either $\tau _{T}<\tau_{C}$, or $\tau_{C}=\infty $ irrespective of whether $\tau _{T}$ is
		finite or not. For most duration distributions considered in the literature, $\tau _{T}=\infty =\tau_{C}$; hence, $\tau _{Y}=\infty $. When $\tau_{C}<\tau _{T}$, $F\neq F_{\ast }$ in general, and $F$ cannot be consistently estimated as $T$ is not observed beyond $\tau _{C}$. When $\tau_{T}=\tau _{C}<\infty ,$ $F=F_{\ast }$ depending on the local structure of $F_{T}$ and $F_{C}$ around the common endpoint. Notice that the above endpoint conditions for $F=F_{\ast }$ can be satisfied with discontinuous $F$. }
\end{remark}

In order to identify $\theta_{0}(t)$, we ensure that $F=F_{\ast }$ by imposing the following condition; see \cite{Stute1999} for a similar assumption in the case of nonlinear regression with randomly censored outcomes.

\begin{assumption}
	\label{A3} \textnormal{$\tau _{T}<\tau _{C}$ or $\tau _{C}=\infty .$}
\end{assumption}
Thus, $\theta _{0}\left( t\right) $ is identified as the parameter value that maximizes $Q\left( \theta, t \right) $ under standard conditions in binary regression, e.g., \citet[Section 9.2.2]{Amemiya1985}, or more recently, \citet[Example 5.40]{VanderVaart1998}.

Henceforth, $Y_{1:n}\leq ...\leq Y_{n:n}$ are the ordered $Y-values$, where
ties within outcomes $T$ or within censoring times $C$ are ordered arbitrarily and
ties among $T$ and $C$ are treated as if the former
precedes the latter. For observations $\left\{ \xi _{i}\right\} _{i=1}^{n}$
of a random variable $\xi $, which may be $\delta $ or $X$, $\xi _{\left[ i:n	\right] }$ is the $i-th$ $\xi -$concomitant of the order statistics $\left\{Y_{i:n}\right\} _{i=1}^{n},$ i.e., $\xi _{\left[ i:n\right] }=\xi _{j}$ if $Y_{i:n}=Y_{j}$.

In the absence of covariates, $F_{\ast T}(t)=F_{\ast }\left( t,\infty
,...,\infty \right) $ is consistently estimated by the \cite{Kaplan1958}
product-limit estimator,
\begin{equation*}
	\hat{F}_{Tn}(t)=1-\prod_{Y_{i:n}\leq t}\left( 1-\frac{\delta _{\left[ i:n%
			\right] }}{n-i+1}\right) ,\text{ }t\in \mathbb{R}^{+}.
\end{equation*}%
By noticing that the jumps of $\hat{F}_{Tn}$ at $Y_{i:n}$ are 
\begin{equation}
	W_{in}=\hat{F}_{Tn}(Y_{i:n})-\hat{F}_{Tn}(Y_{i-1:n})=\frac{\delta _{\left[
			i:n\right] }}{n-i+1}\prod_{j=1}^{i-1}\left( 1-\frac{\delta _{\left[ j:n%
			\right] }}{n-j+1}\right) ,  \label{KMweights}
\end{equation}%
we can express $\hat{F}_{Tn}$ in an additive form as 
\begin{equation*}
	\hat{F}_{Tn}(t)=\sum_{i=1}^{n}W_{in}\cdot 1_{\left\{ Y_{i:n}\leq t\right\} }.
\end{equation*}%

When covariates are present, the natural $F_{\ast }\left( t,x\right) $
estimator is 
\begin{equation}
	\hat{F}_{n}(t,x)=\sum_{i=1}^{n}W_{in}\cdot 1_{\left\{ Y_{i:n}\leq t,X_{\left[
			i:n\right] }\leq x\right\} },  \label{KMmulti}
\end{equation}%
see, \cite{Stute1993, Stute1996a}.  \cite{Stute1993} showed that, under Assumptions \ref{A1}-\ref{A3},
\begin{equation*}
	\lim_{n\rightarrow \infty }\sup_{\left( t,x\right) \in \mathcal{T}\times \mathbb{R}%
		^{k}}\left\vert \left( \hat{F}_{n}-F\right) \left( t,x\right) \right\vert
	=0~a.s.
\end{equation*}

Thus, (\ref{KMmulti}) is indeed a natural candidate to replace $\tilde{F}_{n}$ in (%
\ref{11}) under censoring. This suggests estimating $Q_{\ast }\left(
t,\theta \right) $ by the Kaplan-Meier integral,
\begin{equation*}
	\hat{Q}_{n}\left( \theta ,t\right) =\int \ln p_{\theta }(1_{\left\{ y\leq
		t\right\} },x)\hat{F}_{n}(dy,dx)=\sum_{i=1}^{n}W_{in}\cdot \ln p_{\theta
	}(1_{\left\{ Y_{i:n}\leq t\right\} },X_{\left[ i:n\right] }),
\end{equation*}%
which is a weighted version of $\tilde{Q}_{n}\left( \theta ,t\right)$. 

The Kaplan-Meier distribution regression (KMDR) estimator of $\theta _{0}(t)$ is then given by%
\begin{equation*}
	\hat{\theta}_{n}\left( t\right) =\underset{\theta \in \Theta }{\arg \max }%
	\text{ }\hat{Q}_{n}\left( \theta ,t\right) .
\end{equation*}%
Notice that, to compute the KMDR estimators, one simply needs to run a weighted binary regression, where the weights are the (random) Kaplan-Meier weights $W_{in}$. In the absence of censoring, i.e., $Y_{i}=T_{i},$ $\delta _{i}=1$, $i=1,..,n$, we have that the Kaplan-Meier weights $W_{in}=n^{-1}$, and $\hat{\theta}%
_{n}\left( t\right) $ reduces to $\tilde{\theta}_{n}\left( t\right) .$

\section{Asymptotic Theory}\label{Asy}
\setcounter{equation}{0}

In this section, we present the asymptotic properties of our proposed KMDR
estimator $\hat{\theta}_{n}(t)$ for $\theta _{0}\left( t\right)$.

In order to  establish consistency, we impose the following fairly weak assumption which is compatible with discrete, continuous and mixed duration outcomes.

\begin{assumption}
	\label{A4} The CDF specification $(\ref{dr})$ holds, where $\Lambda :\mathbb{R\mapsto }\left[
	0,1\right] $\ is a continuously differentiable monotone function, the
	distribution of $X$ is not concentrated on an affine subspace of $\mathbb{R}%
	^{k-1},$ and $\mathbb{E}\left\Vert X\right\Vert ^{2}<\infty .$
\end{assumption}

Next theorem, like any other result in the paper, is proved in the Supplemental Appendix. The proof applies the arguments in Example 5.40 in \cite{VanderVaart1998} to show
that 
\begin{gather*}
	\sup_{\theta \in \Theta }\left\vert \left( \hat{Q}_{n}-Q\right) \left(
	\theta ,t\right) \right\vert =o_{p}(1), \\
	\sup_{\theta :\left\Vert \theta -\theta _{\ast }(t)\right\Vert \geq
		\varepsilon }Q\left( \theta ,t\right) <Q\left( \theta _{0}(t),t\right) \text{
		for all }\varepsilon >0\text{ and each }t\in \mathcal{T}\text{.}
\end{gather*}

\begin{theorem}
	\label{Th1Cens}Under Assumptions \ref{A1}, \ref{A2}, \ref{A3} and \ref{A4}, any sequence of estimators $\left\{ \hat{\theta}_{n}(t)\right\} _{n\geq 1}$ that satisfies 
	$$\hat{Q}_{n}\left( \hat{\theta}_{n}(t)\right) \geq \hat{Q}_{n}\left( \theta
	_{0}(t)\right) -o_{p}\left( 1\right),$$  converges in probability to $\theta _{0}(t)$, as $n \rightarrow \infty$.
\end{theorem}

In order to provide the finite dimensional asymptotic distribution of $\hat{\theta}_{n}(t),$ we need the following assumption.

\begin{assumption}
	\label{A5} $\theta _{0}\left( t\right) $ is an inner point of $\Theta $,
	which is a compact subset of $\mathbb{R}^{1+k}.$ Also, let $\mathcal{V} = \{\boldsymbol{x}^{\prime }\theta : \boldsymbol{x}=(1,x^{\prime })^{\prime },  x \in \mathcal{X}, \theta \in \Theta \} $; for $v \in \mathcal{V}$, $\Lambda \left(v \right) $ is bounded
	away from zero and one, and admits a continuously differentiable Lebesgue
	density $\lambda (v)$.
\end{assumption}

The restriction on $\Lambda$ in Assumption  \ref{A5} is a classical assumption, which can be found, for instance, in \citet{Amemiya1985} Assumption 9.2.1. The assumption is satisfied by the normal and logistic distribution (probit and logit), but also for many other distributions. In the general case, Assumption \ref{A5} essentially rules out extremes $t$, since, in these cases, $\Lambda$ may be near zero or one. This assumption can be relaxed at the cost of more involved proofs.

The score function is given by
\begin{equation*}
	\hat{\Psi}_{n}\left( \theta ,t\right) =\frac{\partial }{\partial \theta }%
	\hat{Q}_{n}\left( \theta ,t\right) =\sum_{i=1}^{n} W_{ni} \cdot \psi _{\theta
		t}\left( Y_{i:n},X_{\left[ i:n\right] }\right),
\end{equation*}%
where
\begin{equation*}
	\psi _{\theta t}\left( y,x\right) =\frac{1_{\left\{ y\leq t\right\}
		}-\Lambda (\mathbf{x}^{\prime }\theta )}{\Lambda (\mathbf{x}^{\prime }\theta
		)\left[ 1-\Lambda (\mathbf{x}^{\prime }\theta )\right] }\lambda (\mathbf{x}%
	^{\prime }\theta )\mathbf{x}.
\end{equation*}%
The conditional Fisher information that the binary variable $1_{\left\{
	T\leq t\right\} }$ contains about $\theta _{0}(t)$ given $X$ is $\mathcal{I}_{0}\left( t\right) =\mathcal{I}\left( \theta _{0}\left( t\right) ,t\right)$, where%
\begin{equation}
	\mathcal{I}\left( \theta ,t\right) =\int_{\mathbb{R}^{1+k}}\psi _{\theta
		t}\left( y,x\right) \psi _{\theta t}^{\prime }\left( y,x\right) F\left(
	dy,dx\right)\label{var1}.
\end{equation}%
Notice that, under our assumptions, $\theta \mapsto \mathcal{I}\left( \theta ,\cdot \right) $ is
continuously differentiable.

We first pay attention to the asymptotic distribution of the score $\hat{\Psi%
}_{n}^{0}\left( t\right) =\hat{\Psi}_{n}\left( \theta _{0}(t),t\right) .$
Applying \cite{Stute1995a, Stute1996a} lemmata, $\hat{\Psi}_{n}\left( \theta ,t\right) $
can be expressed as an $U-$statistic of order three with H\'{a}jek
projection $\hat{U}_{n}^{0}(t)=\hat{U}_{n}(\theta _{0}(t),t)$, where%
\begin{equation*}
	\hat{U}_{n}(\theta ,t)=\frac{1}{n}\sum_{i=1}^{n}\zeta _{\psi _{\theta
			t}}(Y_{i},X_{i},\delta _{i}),
\end{equation*}%
and for any function $\left( y,x\right) \mapsto \varphi \left( y,x\right) $,
\begin{equation}
	\zeta _{\varphi }(y,x,d)=\varphi (y,x)\gamma ^{(0)}(y)d+\gamma _{\varphi
	}^{\left( 1\right) }(y)(1-d)-\gamma _{\varphi }^{(2)}(y),  \label{lin}
\end{equation}%
\begin{eqnarray}
	\gamma ^{(0)}(y) &=&\exp \left\{ \int_{-\infty }^{y-}\frac{F_{Y}^{0}(dw)}{%
		1-F_{Y}(w)}\right\} ,  \label{gam0} \\
	\gamma _{\varphi }^{\left( 1\right) }(y) &=&\frac{1}{1-F_{Y}(y)}\int
	1_{\left\{ y<w\right\} }\varphi (w,x){\gamma }^{(0)}(w)F^{1}(dw,dx),
	\label{gam1} \\
	\gamma _{\varphi }^{\left( 2\right) }(y) &=&\int \int \frac{1_{\left\{
			v<y,v<w\right\} }\varphi (w,x){\gamma }^{(0)}(w)}{\left[ 1-F_{Y}(v)%
		\right] ^{2}}F_{Y}^{0}(dv)F^{1}(dw,dx),  \label{gam2} 
\end{eqnarray}
where $F_{Y}^{0}(t) =\mathbb{P}\left( Y\leq t,\delta =0\right) \text{ and }
F^{1}(t,x)=\mathbb{P}\left( Y\leq t,X\leq x,\delta =1\right)$.

To derive the asymptotic normality of the score, we need the following extra moment conditions.
\begin{assumption}
	\label{A6}{For each }$t\in 
	\mathcal{T}$, 
	\begin{align*}
		\left( a\right) \text{ }\mathbb{E}\left[ \left\Vert \psi _{\theta
			_{0}(t)t}(Y,X)\gamma ^{\left( 0\right) }(Y)\delta \right\Vert ^{2} \right] &
		<\infty \\
		\left( b\right) \text{ }\int  \left\Vert \psi _{\theta
			_{0}(t)t}(y,x)\right\Vert \sqrt{S(y)} F(dy,dx) & <\infty ,
	\end{align*}
with
	\begin{equation*}
		S\left( y\right) =\int_{0}^{y-}\frac{F_{C}\left( d\bar{y}\right) }{\left[
			1-F_{Y}\left( \bar{y}\right) \right] \left[ 1-F_{C}\left( \bar{y}\right) %
			\right] }.
	\end{equation*}
\end{assumption}

Assumption \ref{A6}$\left( a\right)$ guarantees that the variance of the leading term in $\zeta _{\psi _{\theta _{0}\left( t\right) t}}$ is bounded, and implies that  $\mathbb{E}\left\Vert \zeta _{\psi _{\theta t}}(Y,X,\delta )\right\Vert^{2}$ is finite. The bias of the Kaplan-Meier integral is not necessarily $o\left( n^{-1/2}\right) $ for any integrand function, and may decrease to
zero at a polynomial rate depending on the degree of censoring, which is
characterized by the function $S$. Assumption \ref{A6}$\left( b\right)$ on $\psi _{\theta_{0}(t)t}$ guarantees that the $\hat{\Psi}_{n}^{0}\left( t\right) $ bias is
of order $o\left( n^{-1/2}\right) $; see \cite{Stute1994a}, \cite{Chen1996}, and \cite{Chen1997}.

Let $\left\{ Z\left( t\right) \right\}_{t=t_{1}}^{t_{m}}$ be a $k+1$ Gaussian random vector with zero mean and covariances%

\begin{equation}
	Cov\left( Z\left( t_{j}\right) Z\left( t_{m}\right) \right) =\mathbb{E}\left[\zeta _{\psi _{\theta _{0}\left( t_j\right) t_j}}(Y,X,\delta )\zeta _{\psi _{\theta _{0}\left( t_m\right) t_m}}^{\prime
	}(Y,X,\delta )\right]=\Omega _{0}\left(j,m\right), \label{Sigma_cov}
\end{equation}

with $j,m=1,...,k+1$.

With Assumption \ref{A6}, we show that $\hat{\Psi}_{n}^{0}\left(
t\right) $ is asymptotically equivalent to a $U-statistic$ of order 3 with
H\'{a}jek projection $\hat{U}_{n}^{0}\left( t\right)$. Thus, $\sqrt{n}\hat{U}_{n}^{0}$ is asymptotically normal with finite dimensional covariance $\Omega _{0}\left(j,m\right)$.  Theorem \ref{Th4.2} below follows by applying Theorem 5.21 in \cite{VanderVaart1998}.

\begin{theorem}\label{Th4.2}
	Under Assumptions \ref{A1}, \ref{A2}, \ref{A3}, \ref{A4}, \ref{A5}, and \ref{A6}, for each $t\in \mathcal{T}$,%
	\begin{equation}
		\hat{\theta}_{n}\left( t\right) =\theta _{0}\left( t\right) +\mathcal{I}%
		_{0}^{-1}\left( t\right) \cdot \hat{U}_{n}^{0}\left( t\right) +o_{p}\left(
		n^{-1/2}\right)  \label{a1}
	\end{equation}%
	and%
	\begin{equation*}
		\left\{ \sqrt{n}\left( \hat{\theta}_{n}-\theta _{0}\right) \left( t\right)
		\right\} _{t=t_{1}}^{t_{m}}\rightarrow _{d}\left\{ \mathcal{I}%
		_{0}^{-1}\left( t\right) Z(t)\right\} _{t=t_{1}}^{t_{m}}.
	\end{equation*}
\end{theorem}

In order to conduct inferences on $\theta_0(\cdot)$, $\Omega _{0}\left( j,m\right) $
is estimated by 
\begin{equation*}
	\hat{\Omega}_{n}\left( j,m\right) =\frac{1}{n}\sum_{i=1}^{n}\hat{\zeta}%
	_{i}(t_j)\hat{\zeta}_{i}^{\prime }(t_m),
\end{equation*}%
where
\begin{eqnarray}
	\hat{\zeta}_{i}(t) &=&\hat{\psi}_{i}(t)\hat{\gamma}_{i}^{(0)}\delta _{i}+\hat{\gamma}_{i}^{\left( 1\right) }(t)(1-\delta )-\hat{\gamma}_{i}^{(2)}(t), \label{asylin}	\\
	\hat{\gamma}_{i}^{(0)} &=&\exp \left\{ \frac{1}{n}\sum_{j=1}^{n}1_{\left\{Y_{j}<Y_{i}\right\} }\frac{1-\delta _{j}}{1-\hat{F}_{nY}(Y_{j})}\right\} , \nonumber\\	
	\hat{\gamma}_{i}^{\left( 1\right) }(t) &=&\frac{1}{1-\hat{F}_{Y}(Y_i)}\frac{1}{n}\sum_{j=1}^{n}1_{\left\{ Y_{i}<Y_{j}\right\} }\hat{\psi}_{j}(t)\hat{\gamma}_{j}^{(0)}\delta _{j},\nonumber \\	
	\hat{\gamma}_{i}^{\left( 2\right) }(t) &=&\frac{1}{n^{2}}\sum_{j=1}^{n}\sum_{\ell =1}^{n}\frac{1_{\left\{ Y_{j}<Y_{i},Y_{j}<Y_{\ell }\right\} }\hat{\psi}_{\ell }(t)\hat{\gamma}_{\ell }^{(0)}}{\left[ 1-\hat{F}_{Y}(Y_{j})
		\right] ^{2}}\left( 1-\delta _{j }\right)\delta _{\ell} ,\nonumber \\
	\hat{\psi}_{i}(t) &=&\psi _{\hat{\theta}_{n}(t)t}\left( Y_{i},X_{i}\right) ,%
	\text{ }\hat{F}_{Y}(t)=\frac{1}{n}\sum_{i=1}^{n}1_{\left\{ Y_{i}\leq
		t\right\}, }\nonumber
\end{eqnarray}
and $\mathcal{I}_{0}\left( t\right) $ is estimated by%
\begin{equation}
	\widehat{\mathcal{I}}_{n}\left( t\right) =\sum_{i=1}^{n} W_{ni}\cdot \hat{\psi}_{i:n}(t)\hat{\psi}_{i:n}^{\prime }(t),\label{Fisher}
\end{equation}
where  $\hat{\psi}_{i:n}(t) = \psi _{\hat{\theta}_{n}(t)t}\left( Y_{i:n},X_{[i:n]}\right)$\footnote{Alternatively, one can estimate $-\mathcal{I}_{0}\left( t\right) $ using the Hessian of $\hat{Q}_n(\hat{\theta}_n,t)$. We follow this path in our simulations.}.
\begin{corollary}	\label{Corollary_4.1 } 
Under the assumptions of Theorem \ref{Th4.2}, $\hat{\Omega}_{n}\left(
	j,m\right) \rightarrow _{p}\Omega _{0}\left( j,m\right) $\textit{\ and }$%
	\widehat{\mathcal{I}}_{n}\left( t_j\right) \rightarrow _{p}\mathcal{I}%
	_{0}^{-1}\left( t_j\right) $\textit{\ for all }$t_j,t_m \in \mathcal{T}$.
\end{corollary}

We can also justify inferences with the assistance of a multiplier
bootstrap technique in the spirit of Theorem 2.3 in \cite{Stute2000}
and Theorem 4 in \cite{SantAnna2016a}, using an external resample of $\left\{ \hat{\zeta}_{i}(t)\right\} _{i=1}^{n}.$ Once we generate $iid$
random numbers $\left\{ V_{i}\right\} _{i=1}^{n}$ independently of the data with mean $0$, variance $1$, and finite third moment, the ``resampled'' $\left\{ \hat{\zeta}_{i}^{\ast
}(t)\right\} _{i=1}^{n}$, with $\hat{\zeta}_{i}^{\ast }(t)=\hat{\zeta}%
_{i}(t)V_{i}$, forms a basis to compute the bootstrap analog of $\hat{\theta}_{n}(t)$ based on the asymptotic linearization (\ref{a1}),
\begin{equation}
\label{eq:boot_theta}
	\hat{\theta}_{n}^{\ast }\left( t\right) =\hat{\theta}_{n}(t)+\widehat{%
		\mathcal{I}}_{n}^{-1}\left( t\right) \hat{U}_{n}^{\ast }\left( t\right) ,
\end{equation}%
where%
\begin{equation*}
	\hat{U}_{n}^{\ast }\left( t\right) =\frac{1}{n}\sum_{i=1}^{n}\hat{\zeta}%
	_{i}^{\ast }(t)
\end{equation*}%
is the bootstrap version of $\hat{U}_{n}^{0}\left( t\right)$.

\begin{theorem}
	\label{thm4.3} Let the assumptions of Theorem \ref{Th4.2} hold. Then, for $t_{1},...,t_{m}\in \mathcal{T}$, we have that
	$\left\{ \sqrt{n}\left( \hat{\theta}_{n}^{\ast }-\hat{\theta}_{n}\right)
	\left( t\right) \right\} _{t=t_{1}}^{t_{m}}$ converges
		in distribution under the bootstrap law to $\left\{ \mathcal{I}%
	_{0}^{-1}\left( t\right) Z(t)\right\} _{t=t_{1}}^{t_{m}}$, with probability 1.
\end{theorem}

The theorem follows by checking the conditional version of the Linderberg-Levy
conditions to show that, with probability 1, $\hat{U}_{n}^{\ast }\left(
t\right) $ converges in distribution to $Z$ under the bootstrap law, i.e., conditional on the data, with probability 1.

Next, we strengthen the pointwise results in Theorems \ref{Th4.2} and \ref{thm4.3} to hold uniformly in $t$. Towards this end, let the space $\ell ^{\infty }\left( \mathcal{G}\right) ^{q}$ be the set of
all $q\times 1$ vectors of bounded functions on $\mathcal{G}$, that is, all the
functions' vectors $g:u\mapsto \mathbb{R}$ such that $\sup_{u\in \mathcal{G}}\left\Vert g\left( u\right) \right\Vert <\infty .$ $\mathcal{G}$ can be
either an Euclidean or a functional space. We interpret the multivariate
empirical process indexed by functions in $\mathcal{G}$ as a random element
in the metric space $\ell ^{\infty }\left( \mathcal{G}\right) ^{q}$ endowed with the 
sup-norm. The empirical process $\hat{\Psi}_{n}$ is indexed by $\left(
\theta ,t\right) \in \Theta \times \mathcal{T}$; in this case $\mathcal{G}%
=\Theta \times \mathcal{T}_{0},$ where $\mathcal{T}_{0}$ is the compact
interval of $\mathcal{T}$ of interest. But $\hat{\Psi}_{n}$ can also be
interpreted as an empirical process indexed by the class $\mathcal{F}$ of
functions $\psi _{\theta t}:\mathcal{X\times }\mathcal{T\times }\left[ 0,1%
\right] \mapsto \mathbb{R}^{1+k};$ in this case $\mathcal{G}=\mathcal{F}$.
The random process $\hat{\Psi}_{n}$ is viewed as a random element of the
metric space $\ell ^{\infty }\left( \Theta \times \mathcal{T}_{0}\right)
^{k+1}$ or $\ell ^{\infty }\left( \mathcal{F}\right) ^{k+1},$ where $%
\mathcal{T}_{0} \in \mathcal{T}$ is the compact subset of $\mathcal{T}$ we are interested in.

In order to establish the weak convergence of $\sqrt{n}\left( \hat{\theta}%
_{n}-\theta _{0}\right) $ as a random element of the metric space $\ell
^{\infty }\left( \mathcal{T}_{0}\right) ^{k+1}$, we impose the following additional assumption, which is also assumed in \citet{Chernozhukov2013a}.

\begin{assumption}
	\label{A7} The duration interval of interest $\mathcal{T}_{0}$ is a compact
	subset of $\mathbb{R}^{+},$ and the conditional distribution function $%
	F_{\left. Y\right\vert X}\left( \left. y\right\vert X\right) $
	admits a Lebesgue density $f_{\left. Y\right\vert X}\left( \left.
	y\right\vert X\right) $ that is uniformly bounded and uniformly continuous
	in $\mathcal{T}_{0}$ with probability 1.
\end{assumption}

Notice that $f_{\left. T\right\vert X}\left( \left. t\right\vert X\right) =%
\dot{\theta}_{0}^{\prime }\left( t\right) \cdot \boldsymbol{X\cdot }\lambda
\left( \theta _{0}^{\prime }\left( t\right) \boldsymbol{X}\right) $ a.s.,
where $\dot{\theta}_{0}\left( t\right) =d\theta _{0}\left( t\right) /dt.$
Therefore, assuming that $f_{\left. T\right\vert X}$ is a proper probability
density may not be innocuous. In practice, $\mathcal{T}_0$ should be bounded by the minimum and maximum value of the observed uncensored duration.

Let $\mathbb{Z}$ be a centered tight Gaussian element of $\ell ^{\infty
}\left( \mathcal{T}_{0}\right) ^{k+1}$ with matrix of variance and
covariance functions 
\begin{equation*}
	\mathbb{E}\left[ \mathbb{Z}\left( t_{1}\right) \mathbb{Z}'\left( t_{2}\right)
	\right] = \Omega _{0}\left( t_{1},t_{2}\right)  ,\text{ }t_{1},t_{2}\in 
	\mathcal{T}_{0},
\end{equation*}
with $\Omega _{0}\left( t_{1},t_{2}\right)$ defined in (\ref{Sigma_cov}).
\begin{theorem}
	\label{Thm4.5} Given a compact subset $\mathcal{T}_{0}$ of $\mathcal{T},$ under Assumptions \ref{A1}, \ref{A2}, \ref{A3}, \ref{A4}, \ref{A5}, \ref{A6}, and \ref{A7},
	$\left\{ \sqrt{n}\left( \hat{\theta}_{n}-\theta _{0}\right) \left( t\right)
	\right\} _{t\in \mathcal{T}_{0}}$ converges weakly to $
	\left\{ \mathcal{I}_{0}^{-1}\left( t\right) \mathbb{Z}\left( t\right)
	\right\} _{t\in \mathcal{T}_{0}}$\textit{\ in }$\ell ^{\infty }\left( 
	\mathcal{T}_{0}\right)^{k+1}.$
\end{theorem}

This functional CLT is proved extending \cite{Chernozhukov2013a} (CFM henceforth) strategy to our setup. To this
end, we first show, using \cite{Stute1995a, Stute1996a} lemmata, that the score function $\hat{\Psi}_{n}$ and $U_n$ are asymptotically equivalent in the metric space $\ell^{\infty}(\Theta, \mathcal{T}_0)^{k+1}$, i.e., 
 $$\sup_{\theta \in \Theta ,t\in \mathcal{T}_{0}}\left\Vert \hat{\Psi}_{n}\left( \theta ,t\right) -U_{n}\left( \theta ,t\right) \right\Vert=o_{p}\left( n^{-1/2}\right), $$ where $U_{n}$ is a $U-process$ with H\'{a}jek
projection $\hat{U}_{n}$. Then, we show that $$\sup_{\theta \in \Theta ,t\in 	\mathcal{T}_{0}}\left\Vert U_{n}\left( \theta ,t\right) -\hat{U}_{n}\left(\theta ,t\right) \right\Vert =o_{p}\left( n^{-1/2}\right) $$ applying
asymptotic results for U-process in \cite{Arcones1993, Arcones1995}. Since the
class $\mathcal{E}$ of functions $\left( y,x,d\right) \mapsto \zeta _{\psi
	_{\theta t}}\left( y,x,d\right) $ is Donsker, $$\left\{ \sqrt{n}\left( \hat{U}%
_{n}\left( \theta ,t\right) -\mathbb{E}\left[\zeta _{\psi _{\theta t}}\left(
Y,X,\delta \right) \right]\right) :\theta \in \Theta ,t\in \mathcal{T}_{0}\right\} $$
converges weakly in $\ell ^{\infty }\left( \Theta \times \mathcal{T}_{0}\right) ^{k+1}$. Then,
noticing that Assumptions \ref{A4} - \ref{A7}  imply Assumption DR in CFM, the functional $\theta \mapsto
\Psi _{0}\left( \theta ,t\right) $, with $\Psi _{0}\left( \theta ,t\right) =
\mathbb{E}\left[\zeta _{\psi _{\theta t}}\left( Y,X,\delta \right)\right]$, is
continuously differentiable for each $t\in \mathcal{T}_{0},$ with $\left.
\left. d\Psi _{0}\left( \theta ,t\right) \right/ d\theta ^{\prime
}\right\rfloor _{\theta =\theta _{0}\left( t\right) }= - \mathcal{I}_{0}\left(
t\right) ,$ which is invertible. Then, according to CFM's Lemma E.3, 
\begin{equation*}
	\sqrt{n}\left( \hat{\theta}_{n}-\theta _{0}\right) \left( t\right) =
	\mathcal{I}_{0}^{-1}\left( t\right) \cdot \hat{\Psi}_{n}\left( \theta
	_{0}\left( t\right) ,t\right) +\hat{r}_{n}\left( t\right) ,
\end{equation*}%
with $\sup_{t\in \mathcal{T}_{0}}\left\Vert \hat{r}_{n}\left( t\right)
\right\Vert =o_{p}\left( 1\right)$. This result and $\mathcal{E}$'s
Donskerness prove the theorem.

A bootstrap version of this theorem follows by applying the multiplier CLT
(see sections 2.6 and 3.6 in \citealp{VanderVaart1996}).

\begin{theorem}
	\label{thm4.6} Suppose the assumptions of Theorem (\ref{Thm4.5}) hold. Then,  with probability 1 $\left\{ 
	\sqrt{n}\left( \hat{\theta}_{n}^{\ast }-\hat{\theta}_{n}\right) \left(
	t\right) \right\} _{t\in \mathcal{T}_{0}}$converges weakly under the bootstrap law to $\left\{\mathcal{I}^{-1}\left( t\right) \mathbb{Z}%
	(t)\right\} _{t\in \mathcal{T}_{0}}$ in $\ell ^{\infty }\left( 	\mathcal{T}_{0}\right)^{k+1} .$ 
\end{theorem}

In next section we discuss applications of these results to test restrictions on $\theta _{0}$, as well as to make inferences on counterfactual distributions and on average distribution marginal effects.

\section{Applications}\label{theory_application}
\setcounter{equation}{0}
\subsection{Testing Linear and Nonlinear Hypothesis}\label{sec:hyp_test}
A natural class of hypotheses to be tested is those of the form%
\begin{equation*}
	H_{0}:\sup_{t\in \mathcal{T}_{0}}\left\Vert \Phi \left( \theta _{0}\left(
	t\right) \right) \right\Vert =0\text{ against }H_{1}:\sup_{t\in \mathcal{T}%
		_{0}}\left\Vert \Phi \left( \theta _{0}\left( t\right) \right) \right\Vert
	> 0,
\end{equation*}%
for some compact subset $\mathcal{T}_{0}$ of $\mathcal{T},$ where $\theta
\mapsto \Phi \left( \theta \right) \in \mathbb{R}^{q}$, $q\leq k+1,$ is a
continuously differentiable map with derivatives $\dot{\Phi}\left( \theta
\right) =\partial \Phi \left( \theta \right) /\partial \theta ^{\prime },$
such that $rank\left( \dot{\Phi}\left( \theta \right) \right) =q$ for all $%
\theta $ in  a neighborhood of $\theta _{0}\left( t\right)$. For instance, when $\Phi \left( \theta\right) = \theta$, the null and alternative reads 
\begin{equation*}
	H_{0}:\sup_{t\in \mathcal{T}_{0}}\left\Vert  \theta _{0}\left(
	t\right)  \right\Vert =0\text{ against }H_{1}:\sup_{t\in \mathcal{T}%
		_{0}}\left\Vert  \theta _{0}\left( t\right)  \right\Vert
	> 0,
\end{equation*}%
which is a significance test for varying coefficients. We can also test that a linear combination of coefficients is satisfied, e.g., $\Phi(\theta) = R \theta - r$, where $R$ and $r$ are known and $rank(R)=k+1$.

A natural
statistic is,%
\begin{equation*}
	\hat{\omega}_{n}=n\cdot \sup_{t\in \mathcal{T}_{0}}\left\Vert \Phi \left( 
	\hat{\theta}_{n}\left( t\right) \right) \right\Vert ^{2}.
\end{equation*}%

By the delta-method, and applying Theorem \ref{Thm4.5},  uniformly in $t\in 
\mathcal{T}_{0},$%
\begin{equation*}
	\sqrt{n}\left[ \Phi \left( \hat{\theta}_{n}\left( t\right) \right) -\Phi
	\left( \theta _{0}\left( t\right) \right) \right] =\dot{\Phi}\left( \theta
	_{0}\left( t\right) \right) \sqrt{n}\left( \hat{\theta}_{n}-\theta
	_{0}\right) \left( t\right)+o_{p}(1)
\end{equation*}%
and under $H_{0},$%
\begin{equation*}
	\hat{\omega}_{n}\rightarrow _{d}\omega _{\infty }=\sup_{t\in \mathcal{T}%
		_{0}}\left\Vert \dot{\Phi}\left( \theta _{0}\left( t\right) \right) \mathcal{%
		I}_{0}^{-1}\left( t\right) \mathbb{Z}\left( t\right) \right\Vert ^{2}.
\end{equation*}%
Since the $\omega _{\infty }^{\prime }s$ distribution depends on $\theta
_{0}\left( \cdot \right) $ and other unknown features of the underlying data
generating process, analytical critical values seems unfeasible, at least with this level of generality. However, critical values can be estimated using the bootstrapped estimator of $\theta _{0}\left( t\right)$

Applying Theorem \ref{thm4.6}, uniformly in $t\in \mathcal{T}_{0},$ for almost every
sample, it follows that
\begin{equation*}
	\sqrt{n}\left[ \Phi \left( \hat{\theta}_{n}^{\ast }\left( t\right) \right) 
	\mathcal{-}\Phi \left( \hat{\theta}_{n}\left( t\right) \right) \right] =\dot{%
		\Phi}\left( \hat{\theta}_{n}\left( t\right) \right) \sqrt{n}\left( \hat{%
		\theta}_{n}^{\ast }-\hat{\theta}_{n}\right) \left( t\right) +o_p(1).
\end{equation*}%
Therefore, we can use the bootstrap test statistic 
\begin{equation}
\label{eq:boot_test_stat}
	\hat{\omega}_{n}^{\ast }=n\cdot \sup_{t\in \mathcal{T}_{0}}\left\Vert \Phi
	\left( \hat{\theta}_{n}^{\ast }\left( t\right) \right) \mathcal{-}\Phi
	\left( \hat{\theta}_{n}\left( t\right) \right) \right\Vert ^{2}.
\end{equation}%

By Theorem \ref{thm4.6}, under $H_{0}$, with probability 1, $\hat{\omega}_{n}^{\ast }$ converges in distribution under the bootstrap law to $\omega _{\infty }$ . Under $H_{1},$ $\hat{\omega}_{n}^{\ast }=O\left( 1\right) $ with probability
1.

We now describe a practical bootstrap algorithm to conduct such types of tests. 

\begin{algorithm}[Bootstrapped-based hypothesis testing]
\label{alg:boot_test}

~\vspace{-.1cm} \setcounter{bean}{0}

\begin{list}
{\textsc{Step} \arabic{bean}.}{\usecounter{bean}}
\item \textnormal{Generate $iid$ $\{V_{i}\}_{i=1}^{n}$ from a distribution with mean 0, variance 1 and finite third moment, e.g., Rademacher distribution.}

\item \textnormal{For a grid $t=t_{1},\dots ,t_{p}$, compute $\left\{ \hat{\theta}_{n}^{\ast }\left( t\right) \right\} _{t=t_{1}}^{t_{p}}$%
as in (\ref{eq:boot_theta}), using the same $\{V_{i}\}_{i=1}^{n}$ for all $t$'s.}

\item  \textnormal{Compute the bootstrap test statistics $\hat{\omega}_{n}^{\ast}$ in (\ref{eq:boot_test_stat})}

\item  \textnormal{Repeat steps 1-3 $B$ times.}

\item \textnormal{Reject $H_{0}$ at the $\alpha -level$ of significance if either} $\hat{\omega}_{n}>\hat{\omega}_{n,\alpha }^{\ast }$, \textnormal{where} $\hat{\omega}_{n,\alpha }^{\ast }$ \textnormal{is the empirical $(1-\alpha)$-quantile of the $B$ bootstrap draws of} $\hat{\omega}_{n}^{\ast}$\textnormal{, or if p-val$^{*} > \alpha$, where $$p\text{-}val^{*} = \frac{1}{B}\sum_{j=1}^B 1_{\left\{\hat{\omega}_{n}^{\ast (j)} > \hat{\omega}_{n} \right\}.}$$}

\end{list}
\end{algorithm}

An interesting application of this type of test is to testing constancy of
some functional of $\theta _{0}$.  For instance,  one can set $$\Phi \left( \theta \right) =\left[ \boldsymbol{0}\vdots I_{K}%
\right] \theta$$ to test that all the slope coefficients are constant.  A
test statistic for this hypothesis is
\begin{equation*}
	\tilde{\omega}_{n}=n \cdot\sup_{t\in \mathcal{T}_{0}}\left\Vert \Phi \left( \hat{%
		\theta}_{n}\left( t\right) \right) - \bar{\hat{\Phi}}_n \right\Vert ^{2},
\end{equation*}%
with $\bar{\hat{\Phi}}_n = \int_{s\in \mathcal{T}_0} {\Phi \left( \hat{\theta}_{n}\left( s\right) \right)} \Psi(ds)/\int_{s\in \mathcal{T}_0} \Psi(ds)$, $\Psi$ being a known probability measure.\footnote{In practice, one can set $\Psi$ to be equal to the empirical distribution of the censored outcome $Y$, though, in this case, the bootstrap procedure needs to be adjusted to account for this additional source of randomness. See Section \ref{sec:adme} for related results.}
Bootstrap critical values  can be
obtained using the bootstrapped statistic,
\begin{equation*}
	\tilde{\omega}_{n}^{\ast }=  n \cdot \sup_{t\in \mathcal{T}_{0}}\left\Vert \left[ \Phi
	\left( \hat{\theta}_{n}^{\ast }\left( t\right) \right) -\bar{\hat{\Phi}}_n^* \right] \mathcal{-}\left[ \Phi \left( \hat{\theta}_{n}\left( t\right)
	\right) - \bar{\hat{\Phi}}_n \right] \right\Vert ^{2},
\end{equation*}
with $\bar{\hat{\Phi}}_n^* =  \int_{s\in \mathcal{T}_0} {\Phi \left( \hat{\theta}_{n}^*\left( s\right) \right)} \Psi(ds)/\int_{s\in \mathcal{T}_0} \Psi(ds)$.

\subsection{Inferences on $F_{\left. T\right\vert X}$}

Theorems \ref{Thm4.5} and \ref{thm4.6} can also be applied to make inferences on $F_{\left.T\right\vert X}\left( \left. t\right\vert x\right) $. The DR $F_{\left.
	T\right\vert X}\left( \left. t\right\vert x\right) ^{\prime }s$ estimator is 
\begin{equation*}
	\hat{F}_{n\left. T\right\vert X}\left( \left. t\right\vert x\right) =\Lambda
	\left( \boldsymbol{x}^{\prime } \hat{\theta}_{n}\left( t\right) \right) ,
\end{equation*}%
with bootstrap analog,%
\begin{equation*}
	\hat{F}_{n\left. T\right\vert X}^{\ast }\left( \left. t\right\vert x\right)
	=\Lambda \left(\boldsymbol{x}^{\prime } \hat{\theta}_{n}^{\ast}\left( t\right) 
	\right) .
\end{equation*}

The asymptotic distribution can be obtained using the delta-method.
Applying Theorem  \ref{Thm4.5}, we have that
\begin{equation*}
	\left\{ \sqrt{n}\left( \hat{F}_{n\left. T\right\vert X}-F_{\left.
		T\right\vert X}\right) \left( \left. t\right\vert x\right) \right\} _{\left(
		t,x\right) \in \mathcal{A}_{0}}\rightarrow _{d}\left\{ \lambda \left( \theta
	_{0}^{\prime }\left( t\right) x\right) \cdot x\cdot \mathcal{I}_{0}^{-1}\left( t\right)\mathbb{Z}\left(
	t\right) \right\} _{\left( t,x\right) \in \mathcal{A}_{0}},
\end{equation*}%
where $\mathcal{A}_{0}=\mathcal{T}_{0}\times \mathcal{X}_{0}$ is a compact
subset of $\mathcal{T}_{0}\times \mathcal{X}_{0}$. Likewise, applying
Theorem \ref{thm4.6}, with probability 1,
\begin{small}
\begin{equation*}
	\left\{ \sqrt{n}\left( \hat{F}_{n\left. T\right\vert X}^{\ast }-\hat{F}%
	_{n\left. T\right\vert X}\right) \left( \left. t\right\vert x\right)
	\right\} _{\left( t,x\right) \in \mathcal{A}_{0}}\rightarrow _{d}\left\{
	\lambda \left( \theta _{0}^{\prime }\left( t\right) x\right) \cdot x\cdot 
	\mathcal{I}_{0}^{-1}\left( t\right)\mathbb{Z}\left( t\right) \right\} _{\left( t,x\right) \in \mathcal{A}_{0}}%
	\text{ in }\ell ^{\infty }\left( \mathcal{A}_{0}\right) .
\end{equation*}
\end{small}

Given two samples $\{Y_i^{(j)}, \delta_i^{(j)}, X_i^{(j)} \}_{i=1}^{n_{j}},~ j=1,2$, under some standard overlap conditions, we can use the conditional CDF estimator computed using sample 1, $\widehat{F}^{(1)}_{T|X,n}$, to compute the marginal counterfactual CDF of population 1 with respect to population 2, $$\widehat{F}^{(1,2)}_{T,n} (t) = \int_\mathcal{X} \widehat{F}^{(1)}_{T|X,n}(t|x) \Tilde{F}^{(2)}_{X,n}(dx) = \dfrac{1}{n_2}\sum_{i=1}^{n_2} \left( \widehat{F}^{(1)}_{T|X,n}(t|X_i^{(2)}\right).$$

\subsection{Average Distribution Marginal Effects }\label{sec:adme}

Although $\hat{\theta}_{n}\left( t\right) $ provides useful information
about the direction/sign of the effect of changes in $X$ on $F_{T|X}\left(
t|X\right) ,$ in general, $\hat{\theta}_{n}(t)$ may not have a clear
economic interpretation. In this section, we argue that this potential
limitation can be easily avoided by focusing on the average distribution
marginal effects (ADME) of $X$,\footnote{ 
	To avoid cumbersome notation, we consider the case where the covariates $X$
	are continuous, and enter the distribution regression model linearly. The
	asymptotic validity of all our results do not rely on this simplification.}
\begin{equation}
\eta _{0}\left( t\right) \equiv \mathbb{E}\left[\beta _{0}\left( t\right) \lambda
\left( \boldsymbol{X}^{\prime } \theta _{0}\left( t\right)\right) \right].   \label{ADME}
\end{equation}
The ADME is the distributional analogue of the popular average partial effects. From (\ref{ADME}), it is clear that he natural estimator for  $\eta _{0}\left( t\right)$ is 
\begin{equation*}
	\hat{\eta}_{n}\left( t\right) =\hat{\beta}_{n}\left( t\right) \frac{1}{n}%
\sum_{i=1}^{n}\lambda \left( \boldsymbol{X}_i^{\prime } \hat{\theta}_{n}\left( t\right) 
\right) .
\end{equation*}%

In what follows, we derive the asymptotic properties of $\hat{\eta}_{n}\left( t\right)$.  As in Theorem \ref{Thm4.5}, we first show that, uniformly in $t \in \mathcal{T}_0$,
\begin{equation}
\left( \hat{\eta}_{n}\left( t\right) - {\eta}_{0}\left( t\right) \right)=\frac{1}{n}\sum_{i=1}^{n}\zeta
_{i, ADME}(t)+o_{p}\left(n^{-1/2}\right) ,
\label{adme-linrep}
\end{equation}%
where 
\begin{eqnarray}
\zeta _{i,ADME}(t) &=&\beta _{0}\left( t\right) \cdot \left( \lambda (\mathbf{
	X}_{i}^{\prime}\theta _{0}(t))-\mathbb{E}\left[ \lambda (\mathbf{
	X}^{\prime}\theta _{0}(t))\right] \right) \nonumber \\ 
&&+\mathbb{E}\left[ \lambda (\mathbf{
	X}^{\prime}\theta _{0}(t))\right] \cdot
H\cdot \mathcal{I}_{0}\left( t\right)^{-1}\cdot \zeta _{i}(t) \label{lin_adme}\\
&&+\beta _{0}\left( t\right) \cdot \mathbb{E}\left[ \dot{\lambda} \left( \mathbf{
	X}^{\prime}\theta _{0}(t)\right) \mathbf{X}^{\prime }\right]
\cdot \mathcal{I}_{0}\left( t\right)^{-1}\cdot \zeta _{i}(t) \nonumber
\end{eqnarray}%
with $\zeta _{i}(t)$ is as defined in (\ref{lin}), with $\varphi = \psi _{\theta
	t}$, $\dot{\lambda}\left(
u\right) =d\lambda (u)/du$, $H\equiv \left[ 0_{k},I_{k}\right] $, $0_{k}$ the $%
k\times 1$ vector of zero, and $I_{k}$ the $k$-dimensional identity matrix.

Let $\mathbb{Z}_{ADME}$ be a centered tight Gaussian element of $\ell ^{\infty
}\left( \mathcal{T}_{0}\right) ^{k}$ with matrix of variance and
covariance functions 
\begin{equation*}
	\mathbb{E}\left[ \mathbb{Z}_{ADME}\left( t_{1}\right) \mathbb{Z}_{ADME}'\left( t_{2}\right)\right]=\mathbb{E}\left[ \zeta_{ADME}\left( t_1\right) \zeta_{ADME}^{\prime }\left( t_2 \right)\right]. 
\end{equation*}
Consider the bootstrapped estimator for the ADME 
 \begin{equation}
	\hat{\eta}_{n}^{\ast }\left( t\right) =\hat{\eta}_{n}(t) + \frac{1}{n}\sum_{i=1}^{n}V_i \cdot \hat{\zeta}%
	_{i, ADME}(t)  \label{adme.boot}
\end{equation}%
where $iid$
random numbers $\left\{ V_{i}\right\} _{i=1}^{n}$ are $iid$ random variables with mean $0$ and variance $1$, generated independently of the sample, and 
\begin{eqnarray}
	\hat{\zeta}_{i,ADME}(t) &=& \hat{\beta}_{n}\left( t\right) \cdot \left( \lambda (\boldsymbol{X}_i^{\prime } \hat{\theta}_{n}\left( t\right) )-\frac{1}{n}\sum_{j=1}^{n} \lambda (\boldsymbol{X}_j^{\prime } \hat{\theta}_{n}\left( t\right) )\right) \nonumber \\ 
	&&+\left(\frac{1}{n}\sum_{j=1}^{n} \lambda (\boldsymbol{X}_j^{\prime } \hat{\theta}_{n}\left( t\right) )\right) \cdot
	H\cdot \widehat{\mathcal{I}}_{n}\left( t\right)^{-1}\cdot \hat{\zeta} _{i}(t) \label{lin_adme}\\
	&&+\hat{\beta} _{n}\left( t\right) \cdot \left(\frac{1}{n}\sum_{j=1}^{n}  \dot{\lambda} \left( \boldsymbol{X}_j^{\prime } \hat{\theta}_{n}\left( t\right) \right) \mathbf{X}_{j}^{\prime }\right)
	\cdot \widehat{\mathcal{I}}_{n}\left( t\right)^{-1}\cdot \hat{\zeta} _{i}(t) \nonumber
\end{eqnarray}%
where $\hat{\zeta} _{i}(t)$ is as in (\ref{asylin}) and $\widehat{\mathcal{I}}_{n}\left( t\right)$ is as in (\ref{Fisher}).

\begin{theorem}
\label{Th6.1} Suppose that the conditions of Theorem \ref{Thm4.5} hold.
Then, $$
\left\{ \sqrt{n}\left( \hat{\eta}_{n}\left( t\right) - {\eta}_{0}\left( t\right) \right)\right\} _{t\in \mathcal{T}_{0}} \rightarrow_{d} \left\{ \mathbb{Z}_{ADME}(t)\right\} _{t\in \mathcal{T}_{0}} \text{ in } \ell ^{\infty
}\left( \mathcal{T}_{0}\right) ^{k}.$$
In addition, if follows that $\left\{ 
\sqrt{n}\left( \hat{\eta}_{n}^{\ast }-\hat{\eta}_{n}\right) \left(
t\right) \right\} _{t\in \mathcal{T}_{0}}$ converges weakly under the bootstrap law to $\left\{ \mathbb{Z}_{ADME}(t)\right\} _{t\in \mathcal{T}_{0}}$
in $\ell ^{\infty
}\left( \mathcal{T}_{0}\right) ^{k}$, with probability one.  
\end{theorem}

We now describe a practical bootstrap algorithm to compute simultaneous
confidence intervals for the ADME associated with a given covariate $X_{1}$

\begin{algorithm}[Bootstrapped Simultaneous Confidence Intervals]
\label{alg:boot1}

~\vspace{-.1cm} \setcounter{bean}{0}

\begin{list}
{\textsc{Step} \arabic{bean}.}{\usecounter{bean}}
\item \textnormal{Generate $iid$ $\{V_{i}\}_{i=1}^{n}$ from a distribution with mean 0, variance 1 and finite third moment, e.g., Rademacher distribution.}

\item \textnormal{For a grid $t=t_{1},\dots ,t_{p}$, compute $\left\{ \hat{\eta}_{n}^{\ast }\left( t\right) \right\} _{t=t_{1}}^{t_{p}}$%
as in (\ref{adme.boot}), using the same $\{V_{i}\}_{i=1}^{n}$ for all $t$'s.}

\item  \textnormal{ Let $\widehat{\kappa }$ and $\widehat{\kappa }^{\ast }$ be the
vectorized $\left\{ \hat{\eta}_{n}\left( t\right) \right\} _{t=t_{1}}^{t_{p}}$ and 
$\left\{ \hat{\eta}_{n}^{\ast }\left( t\right) \right\} _{t=t_{1}}^{t_{p}}$%
, respectively, denote their $j$-$th$-element by $\widehat{\kappa }\left(
j\right) $ and $\widehat{\kappa }^{\ast }\left( j\right) $, and compute}
$
\hat{R}^{\ast }\left( j\right) =\left( \widehat{\kappa }^{\ast }-%
\widehat{\kappa }\right) \left( j\right) .
$

\item  \textnormal{Repeat steps 1-3 $B$ times.}

\item \textnormal{For each bootstrap draw, compute $\bar{R}^{^{\ast }}=\max_{j}\left(
\left\vert \widehat{R}^{\ast }\left( j\right) \right\vert \right) .$}

\item \textnormal{Construct $\widehat{c}_{1-\alpha }$ as the empirical $\left( 1-a\right) 
$-quantile of the $B$ bootstrap draws of $\bar{R}^{^{\ast }}$.}

\item \textnormal{Construct the bootstrapped simultaneous confidence band for $\hat{\eta}_{n}\left(
t\right) $, as}
$\widehat{CB}\left( t\right) =\left[ \hat{\eta}_{n}\left( t\right) \pm \widehat{c}_{1-\alpha }\right]$ .
\end{list}
\end{algorithm}

\section{Simulation Studies}\label{simulation}
\setcounter{equation}{0}

In this section, we briefly summarize the simulation results discussed in
Section \ref{MC-appendix} of the Supplementary Appendix. In short, we
compare the finite sample performance of the proposed KMDR estimators with
those based on the proportional hazard (PH) model and on the proportional
odds (PO) model. We consider three different data generating processes
(DGPs): one DGP that satisfies the PH but not the PO assumption, one DGP
that satisfies the PO but not the PH assumption, and one DGP where both PH
and PO assumptions are violated, though it admits a DR specification with
time-varying slope coefficient. We consider different levels of random
censoring, and compare the KMDR, PH and PO models based on the conditional
CDF, $F_{\left. T\right\vert X}$, and the ADME as defined in (\ref{ADME}).
Both functionals have a clear economic interpretation.

Overall, the simulation results highlight that our proposed KMDR estimators
performs nearly as well as the PH and PO model when these models are
correctly specified, since KMDR nests both PH and PO specifications. On the other hand, when PH and
PO models are misspecified, the proposed KMDR estimators
performs better than these other popular specifications. Such gains are
especially noticeable when one is interested in the ADME$\left( t\right) $;
see Table \ref{tab:dgp3}.
\afterpage{
\begin{table}[!ht]
\caption{Simulated finite-sample properties when DGP does not satisfy neither PH nor PO assumptions.}
\label{tab:dgp3}\centering
\begin{adjustbox}{width=1.\textwidth}
\begin{threeparttable}
\begin{tabular}{ccrccccrccccrcccc}
\toprule
\midrule
&  &  & \multicolumn{4}{c}{No Censoring} &  & \multicolumn{4}{c}{10 \%
Censoring} &  & \multicolumn{4}{c}{30\% Censoring} \\ 
\cline{4-7}\cline{9-12}\cline{14-17}
& $n$ &  & $PH$ & $P0$ & $DR_{cll}$ & $DR_{l}$ &  & $PH$ & $P0$ & $DR_{cll}$
& $DR_{l}$ &  & $PH$ & $P0$ & $DR_{cll}$ & $DR_{l}$ \\ \hline
Average & 100 &  & \multicolumn{1}{r}{0.62} & \multicolumn{1}{r}{0.59} & 
\multicolumn{1}{r}{0.35} & \multicolumn{1}{r}{0.65} &  & \multicolumn{1}{r}{
0.63} & \multicolumn{1}{r}{0.72} & \multicolumn{1}{r}{0.36} & 
\multicolumn{1}{r}{0.69} &  & \multicolumn{1}{r}{0.58} & \multicolumn{1}{r}{
0.78} & \multicolumn{1}{r}{0.98} & \multicolumn{1}{r}{1.24} \\ 
absolute bias & 400 &  & \multicolumn{1}{r}{0.85} & \multicolumn{1}{r}{0.69}
& \multicolumn{1}{r}{0.08} & \multicolumn{1}{r}{0.41} &  & 
\multicolumn{1}{r}{0.73} & \multicolumn{1}{r}{0.64} & \multicolumn{1}{r}{0.12
} & \multicolumn{1}{r}{0.45} &  & \multicolumn{1}{r}{0.67} & 
\multicolumn{1}{r}{0.59} & \multicolumn{1}{r}{0.25} & \multicolumn{1}{r}{0.55
} \\ 
for $F\left( t|X=0.5\right) $ & 1600 &  & \multicolumn{1}{r}{0.91} & 
\multicolumn{1}{r}{0.77} & \multicolumn{1}{r}{0.03} & \multicolumn{1}{r}{0.34
} &  & \multicolumn{1}{r}{0.81} & \multicolumn{1}{r}{0.73} & 
\multicolumn{1}{r}{0.03} & \multicolumn{1}{r}{0.34} &  & \multicolumn{1}{r}{
0.61} & \multicolumn{1}{r}{0.62} & \multicolumn{1}{r}{0.15} & 
\multicolumn{1}{r}{0.46} \\ \hline
Average & 100 &  & \multicolumn{1}{r}{4.29} & \multicolumn{1}{r}{4.42} & 
\multicolumn{1}{r}{4.29} & \multicolumn{1}{r}{4.32} &  & \multicolumn{1}{r}{
4.42} & \multicolumn{1}{r}{4.58} & \multicolumn{1}{r}{4.55} & 
\multicolumn{1}{r}{4.57} &  & \multicolumn{1}{r}{4.86} & \multicolumn{1}{r}{
4.97} & \multicolumn{1}{r}{5.36} & \multicolumn{1}{r}{5.35} \\ 
RMSE & 400 &  & \multicolumn{1}{r}{2.34} & \multicolumn{1}{r}{2.41} & 
\multicolumn{1}{r}{2.12} & \multicolumn{1}{r}{2.20} &  & \multicolumn{1}{r}{
2.42} & \multicolumn{1}{r}{2.52} & \multicolumn{1}{r}{2.28} & 
\multicolumn{1}{r}{2.36} &  & \multicolumn{1}{r}{2.57} & \multicolumn{1}{r}{
2.64} & \multicolumn{1}{r}{2.54} & \multicolumn{1}{r}{2.62} \\ 
for $F\left( t|X=0.5\right) $ & 1600 &  & \multicolumn{1}{r}{1.46} & 
\multicolumn{1}{r}{1.48} & \multicolumn{1}{r}{1.08} & \multicolumn{1}{r}{1.19
} &  & \multicolumn{1}{r}{1.41} & \multicolumn{1}{r}{1.47} & 
\multicolumn{1}{r}{1.09} & \multicolumn{1}{r}{1.20} &  & \multicolumn{1}{r}{
1.40} & \multicolumn{1}{r}{1.50} & \multicolumn{1}{r}{1.25} & 
\multicolumn{1}{r}{1.38} \\ \hline
Average & 100 &  & \multicolumn{1}{r}{17.13} & \multicolumn{1}{r}{18.09} & 
\multicolumn{1}{r}{0.29} & \multicolumn{1}{r}{0.70} &  & \multicolumn{1}{r}{
16.50} & \multicolumn{1}{r}{18.31} & \multicolumn{1}{r}{0.56} & 
\multicolumn{1}{r}{0.74} &  & \multicolumn{1}{r}{16.16} & \multicolumn{1}{r}{
19.15} & \multicolumn{1}{r}{3.14} & \multicolumn{1}{r}{2.85} \\ 
absolute bias & 400 &  & \multicolumn{1}{r}{17.44} & \multicolumn{1}{r}{18.42
} & \multicolumn{1}{r}{0.14} & \multicolumn{1}{r}{0.67} &  & 
\multicolumn{1}{r}{16.72} & \multicolumn{1}{r}{18.68} & \multicolumn{1}{r}{
0.50} & \multicolumn{1}{r}{0.58} &  & \multicolumn{1}{r}{16.15} & 
\multicolumn{1}{r}{19.42} & \multicolumn{1}{r}{1.40} & \multicolumn{1}{r}{
1.07} \\ 
for $ADME\left( t\right) $ & 1600 &  & \multicolumn{1}{r}{17.44} & 
\multicolumn{1}{r}{18.53} & \multicolumn{1}{r}{0.07} & \multicolumn{1}{r}{
0.51} &  & \multicolumn{1}{r}{16.78} & \multicolumn{1}{r}{18.72} & 
\multicolumn{1}{r}{0.06} & \multicolumn{1}{r}{0.53} &  & \multicolumn{1}{r}{
16.19} & \multicolumn{1}{r}{19.56} & \multicolumn{1}{r}{0.76} & 
\multicolumn{1}{r}{0.67} \\ \hline
Average & 100 &  & \multicolumn{1}{r}{21.51} & \multicolumn{1}{r}{23.32} & 
\multicolumn{1}{r}{14.69} & \multicolumn{1}{r}{15.10} &  & 
\multicolumn{1}{r}{21.05} & \multicolumn{1}{r}{23.39} & \multicolumn{1}{r}{
16.00} & \multicolumn{1}{r}{16.33} &  & \multicolumn{1}{r}{21.42} & 
\multicolumn{1}{r}{24.29} & \multicolumn{1}{r}{20.60} & \multicolumn{1}{r}{
20.57} \\ 
RMSE & 400 &  & \multicolumn{1}{r}{18.84} & \multicolumn{1}{r}{20.02} & 
\multicolumn{1}{r}{7.09} & \multicolumn{1}{r}{7.37} &  & \multicolumn{1}{r}{
18.21} & \multicolumn{1}{r}{20.31} & \multicolumn{1}{r}{7.46} & 
\multicolumn{1}{r}{7.75} &  & \multicolumn{1}{r}{17.86} & \multicolumn{1}{r}{
21.06} & \multicolumn{1}{r}{9.62} & \multicolumn{1}{r}{9.87} \\ 
for $ADME\left( t\right) $ & 1600 &  & \multicolumn{1}{r}{17.90} & 
\multicolumn{1}{r}{19.04} & \multicolumn{1}{r}{3.56} & \multicolumn{1}{r}{
3.79} &  & \multicolumn{1}{r}{17.26} & \multicolumn{1}{r}{19.23} & 
\multicolumn{1}{r}{3.77} & \multicolumn{1}{r}{4.01} &  & \multicolumn{1}{r}{
16.71} & \multicolumn{1}{r}{20.10} & \multicolumn{1}{r}{4.73} & 
\multicolumn{1}{r}{4.89} \\  \bottomrule
\end{tabular}
\par
\begin{tablenotes}[para,flushleft]
\small{
Note: Simulations based on one thousand Monte Carlo experiments. ``$PH$'' stands for estimators based on the proportional hazard model. ``$PO$'' stands for estimators based on the proportional odds model. ``$DR_{cll}$'' and ``$DR_{l}$'' stand for estimators based on the proposed distribution regression mode with the $cloglog$ and $logit$ link functions, respectively.}
\end{tablenotes}
\end{threeparttable}
\end{adjustbox}
\end{table}
}
When comparing DR specifications with different link functions, we
notice that estimators that use the cloglog link function tend to be more
robust against model misspecifications than those based on the logit link
function. This is perhaps because the cloglog link function is asymmetric and adapts
better to the non-central parts of the distribution where there are many
zeros (left tail) or many ones (right tail). Thus, we recommend favoring the
cloglog specification in detriment of the logit one, though a more formal
discussion about such choices are beyond the scope of this paper.

\section{The effect of unemployment benefits on unemployment duration}\label{Application}
\setcounter{equation}{0}

One of the main concerns of the design of unemployment insurance policies is
their adverse effect on unemployment duration. The prevailing view of the
economics literature is that increasing unemployment insurance (UI) benefits
leads to higher unemployment duration driven by a moral hazard
effect: higher UI increases the agent's reservation wage and reduces the incentive to job
search, see, e.g., \cite{Krueger2002a} and the references therein. Given
that moral hazard leads to a reduction of social welfare, this argument has
been used against increases in UI benefits.

In a seminal paper, \cite{Chetty2008} challenges the traditional view that
the link between unemployment benefits and duration is only because of moral
hazard. He shows, among many other things, that distortions cause by UI on search behavior are mostly due to a \textquotedblleft liquidity effect\textquotedblright . In
simple terms, UI benefits provide cash-in-hand that allows liquidity
constrained agents to equalize the marginal utility of consumption when
employed and unemployed. Such a liquidity effect reduces the pressure to
find a new job, leading to longer unemployment spells. However, in contrast
to the moral hazard effect, the liquidity effect is a socially beneficial
response to the correction of market failures. Thus, if one finds support in
favor of liquidity effects, increases in UI benefits may lead to
improvements in total welfare. 

In this section, we provide additional evidence of the existence of this liquidity effect
by comparing the effect of UI on household that are liquidity constrained with those that are unconstrained. 
In contrast to \cite{Chetty2008}, we do not rely on the Cox proportional hazard model but rather use our proposed KMDR tools. In this context, the Cox hazard model might be too rigid as does not allow for the effect of UI benefits to vary depending on whether a worker is starting their unemployment claim or they have been unemployed for a while. As we argued before, our proposed tools allow for this richer types of heterogeneity.

As in \cite{Chetty2008}, our data comes from the Survey of Income and
Program Participation (SIPP) for the period spanning 1985-2000. Each SIPP
panel surveys households at four-month intervals for two-four years,
collecting information on household and individual characteristics, as well
as employment status. The sample consists of prime-age males who have
experienced job separation and report to be job seekers, are not on
temporary layoff, have at least three months of work history in the survey
and took up unemployment insurance benefits within one month after job loss.
These restrictions leave 4,529 unemployment spells in the sample, 21.3\% of
those being right-censored. Unemployment durations are measured in weeks
while individuals' UI benefits are measured using the two-step imputation
method described in \cite{Chetty2008}. For further details about the data,
see \cite{Chetty2008}.

To analyze the effect of UI benefits on unemployment duration, we estimate the
DR model%
\begin{equation}
F_{\left. T\right\vert UI,Z}\left( \left. t\right\vert UI,Z\right) =\Psi
\left( \alpha (t)+\beta (t)\ln UI+Z^{\prime }\gamma (t)\right) ,  \label{cdr}
\end{equation}%
where $\Psi \left( u\right) =1-\exp \left( -\exp \left( u\right) \right) $
is the complementary log-log link function, $UI$ is the worker's weekly
unemployment insurance benefits, and $Z$ is a vector of controls including
the worker's age, years of education, marital status dummy, logged
pre-unemployment annual wage, and total wealth. To control for local labor
market conditions and systematic differences in risk and performance across
sector types, $Z$ also includes the state average unemployment rate and
dummies for industry. We note that \cite{Chetty2008} considers additional
controls, including state, year and occupation fixed effects, resulting in a
specification with almost 90 unknown parameters. For this reason, we adopt a
more parsimonious specification described above. Let $\theta \left( t\right)
=\left( \alpha (t),\beta (t),\gamma (t)^{^{\prime }}\right) ^{\prime }$, and 
$\mathbf{X=}\left( 1,\ln UI,Z^{\prime }\right) ^{\prime }$.

Our main goal is to understand the effect of changes in UI on the
probability of one finding a job in $t$ weeks, where $t=2,3,\dots 50.$
Although the sign of $\beta \left( t\right) $ indicates the direction of
such a change, its magnitude may not have a straightforward economic
interpretation. Thus, we focus on the ADME$\left( t\right) $ of\ $\ln UI$, 
\begin{equation}
ADME_{\ln UI}\left( t\right) =\mathbb{E}\left[ \frac{\partial F_{\left.
T\right\vert UI,Z}\left( \left. t\right\vert UI,Z\right) }{\partial \ln UI}%
\right] ,  \label{elast}
\end{equation}%
which is easy to interpret. For instance, an estimate of $ADME_{lnUI}(t)$ equal to 0.1 suggests that, on average, raising unemployment benefits by one percent increases the probability of finding a job in $t$ weeks or less by 0.1 percentage point. As discussed in Theorem \ref{DRTrad}, we can estimate $%
ADME_{\ln UI}\left( t\right) $ by%
\begin{equation}
\widehat{ADME}_{n,\ln UI}\left( t\right) =\frac{1}{n}\sum_{i=1}^{n}\hat{\beta%
}_{n}(t)\cdot \psi \left( \mathbf{X}_{i}^{\prime }\hat{\theta}_{n}(t)\right)
,  \label{elast.hat}
\end{equation}%
where $\psi \left( u\right) =d\Psi \left( u\right) /du$, and $\hat{\theta}%
_{n}\left( t\right) =\left( \hat{\alpha}_{n}(t),\hat{\beta}_{n}(t),\hat{%
\gamma}_{n}(t)^{^{\prime }}\right) ^{\prime }$ is the KMDR estimator of $%
\theta \left( t\right) $. When $\Psi $ is the cloglog link function, $\psi \left( u\right) =\exp \left( u\right) \exp \left( -\exp
\left( u\right) \right) $.

In what follows, we multiply the estimators of ADME$_{\ln UI}\left( t\right) 
$ by 10, which is interpreted as the effect of a 10\% increase in UI
benefits. Figure \ref{fig.amde1} reports the estimates for the full sample
(solid line), together with the 90 percent bootstrapped pointwise, and
simultaneous confidence intervals (dark and light shaded area, respectively)
computed using Algorithm \ref{alg:boot1}.

The result reveals interesting effects. On average, a 10\% increase in UI
benefits appears to have no effect on the probability of a worker finding a
job in the first seven weeks of the unemployment spell. Nonetheless, Figure %
\ref{fig.amde1} shows that a change in UI is associated with a reduction of
the probability of finding a job in the first $t=\left\{ 8,\dots ,50\right\} 
$ weeks. Such an effect seems to be monotone until week 18, where a 10\%
increase in UI is associated with a two-percentage-point decrease in the
average probability of finding a job until that week. After week 18, the
effect of an increase in UI benefits on employment probabilities seems to
weaken but remains statistically significant at the 10\% level, except for $%
t=\left\{ 34,35,39\right\} $. Note that the bootstrap simultaneous
confidence interval is slightly wider than the pointwise one. However, it is
important to mention that the bootstrap uniform confidence interval is
designed to contain the entire true path of the ADME$_{\ln UI}\left(
t\right) $ 90\% of the time, which is in sharp contrast to the bootstrap
pointwise confidence interval. This highlights the practical appeal of using
simultaneous instead of pointwise inference procedures to better quantify
the overall uncertainty in the estimation of all ADMEs.
\afterpage{
\begin{figure}
\begin{center}
\begin{adjustbox}{width = 0.65\textwidth}
\includegraphics{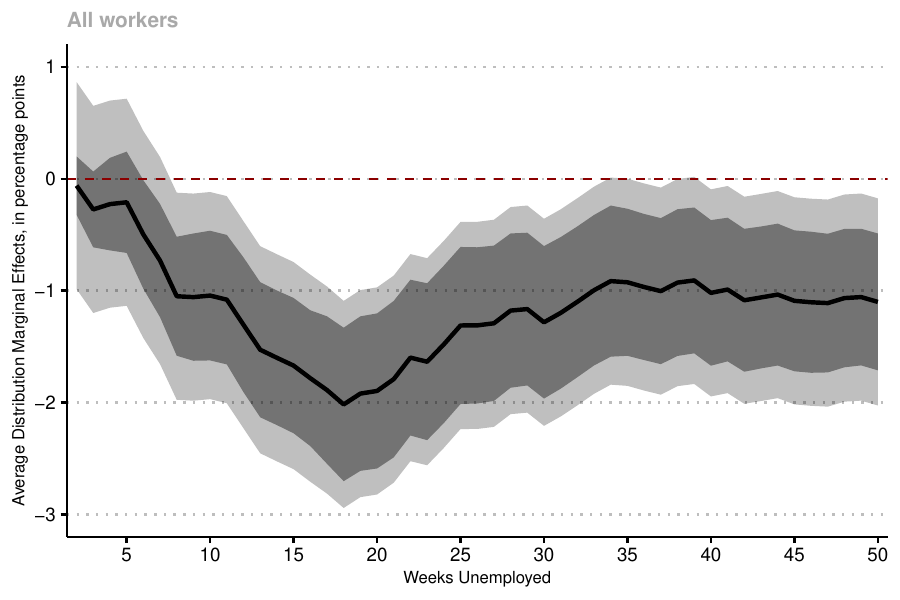}
\end{adjustbox}
\end{center}
\par
\vspace*{-5mm}
\caption{{Estimated average effect of a 10\% increase of unemployment
benefits on the distribution of unemployment duration, full sample. The
solid line represents the estimates while the dark and light shaded areas
are the 90\% bootstrapped based pointwise and simultaneous confidence
interval based on 100,000 bootstrap draws, respectively. Sample size is 4,307.}}
\label{fig.amde1}
\end{figure}
}
Overall, Figure \ref{fig.amde1} shows that although an increase in UI
benefits is close to zero at the beginning of the unemployment spell, they have an
U-shaped effect on the unemployment duration distribution. Interestingly,
estimates of ADME$_{\ln UI}\left( t\right) $ based on the Cox PH model with the same set of covariates as in (\ref{cdr}) suggests that
the effect of UI on unemployment duration distribution is monotone for $t
\in [0,50]$. However, once the proportional hazard specification is tested using either \cite{Grambsch1994} procedure or our bootstrap-based testing procedure for the null hypothesis that all $\hat{\theta}_{n}(t), t=\{2,\dots,50\}$ are constant in $t$, as described in Section \ref{sec:hyp_test}, the null of proportionality is rejected
at the usual significance levels\footnote{The p-value associated with our proposed test statistics is 0.004.}, implying that, indeed, a proportional hazard
model may not be appropriate for this application. Our KMDR model does not
rely on such an assumption.

Although the results in Figure \ref{fig.amde1} show that, on average,
changes in UI have a U-shaped effect on the unemployment duration
distribution, the analysis remains silent about the liquidity effects. To
shed light on the importance of the liquidity effect relative to the moral
hazard effect, \cite{Chetty2008} argues that one can compare the response to
an increase in UI benefits of workers who are not financially constrained
with those who are constrained. Given that unconstrained workers have the
ability to smooth consumption during unemployment, liquidity effects are
absent and UI benefits lengthen unemployment duration only via moral hazard
effects for these subgroup of individuals. To pursue this logic, we follow 
\cite{Chetty2008} and use two proxy measures of liquidity constraint: liquid
net wealth at the time of job loss (\textquotedblleft net
wealth\textquotedblright ) and an indicator for having to make a mortgage
payment. \cite{Chetty2008} argues that workers with higher net wealth are
less sensitive to UI benefit levels because they are less likely to be
financially constrained. Similarly, workers that have to make mortgage
payments before job loss have less ability to smooth consumption during
unemployment because they are unlikely to sell their homes during the
unemployment spell, whereas renters can adjust faster. To assess the role of
liquidity effects, we divide the entire sample into four subsamples: $(a)$
workers with net wealth below the median, $(b)$ workers with net wealth
above the median, $\left( c\right) $ workers with a mortgage, and $\left(
d\right) $ workers without a mortgage. For each subsample, we estimate (\ref%
{cdr}) using the KMDR approach, its corresponding estimator of ADME$_{\ln
UI}\left( t\right) $, $\widehat{ADME}_{n,\ln UI}\left( t\right) $ as in (\ref%
{elast.hat}), and construct 90\% bootstrap pointwise and simultaneous
confidence intervals.

\begin{figure}[!ht!]
\begin{center}
\begin{adjustbox}{width = .8\textwidth}
\includegraphics{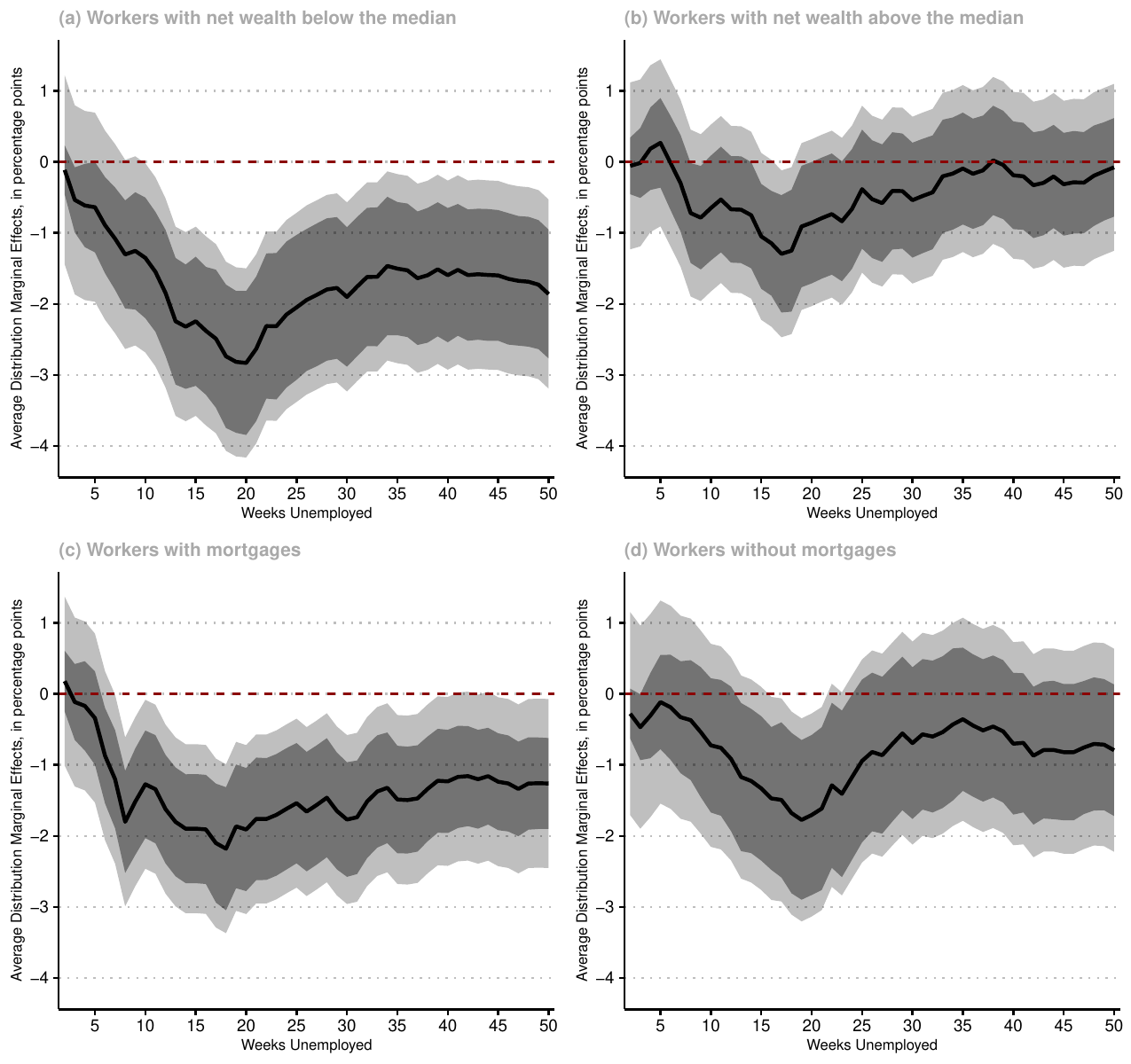}
\end{adjustbox}
\end{center}
\par
\vspace*{-5mm}

\caption{{Estimated average effect of a 10\% increase of unemployment
benefits on the distribution of unemployment duration, in different
subpopulations. The solid line represents the estimates
while the dark and light shaded areas are the 90\% bootstrapped based
pointwise and simultaneous confidence interval based on 100,000 bootstrap
draws, respectively. Sample sizes for Panel (a)-(d) are 2152, 2155, 1952 and 2355, respectively.}}

\label{fig.amde.liquidity}
\end{figure}

Figure \ref{fig.amde.liquidity} reports the average effects of a 10\%
increase in UI benefits on the unemployment duration distribution in each of
the four subsamples. The solid lines are the point estimates and the dark
and light shaded areas are the 90\% bootstrap pointwise and simultaneous
confidence intervals, respectively.

The results show interesting heterogeneity of UI benefits effects with
respect to liquidity constraint proxies. For those workers with net wealth
above the median, an increase in UI benefits has no statistically
significant effect on the unemployment duration distribution, except for $%
t=\left\{ 17,18\right\} $. Thus, for those workers who are not constrained
(and, therefore, for whom the liquidity effect is approximately zero), the
moral hazard effect seems to be close to zero. On the other hand, for those
workers with net wealth below the median, an increase in UI benefits is
associated with lower probabilities of finding a job. The conclusion using
mortgage as a proxy for liquidity constraint is qualitatively the same. Note
that our analysis suggests that the ADME of an increase of UI is
non-monotone across the unemployment duration distribution, highlighting the
flexibility of the DR approach. Indeed, using our proposed test for all slope coefficients being constant, the proportional hazard
specification is rejected at the 10\% significance levels in each subsample.

\begin{figure}[!ht!]
\begin{center}
\begin{adjustbox}{width = .8\textwidth}
\includegraphics{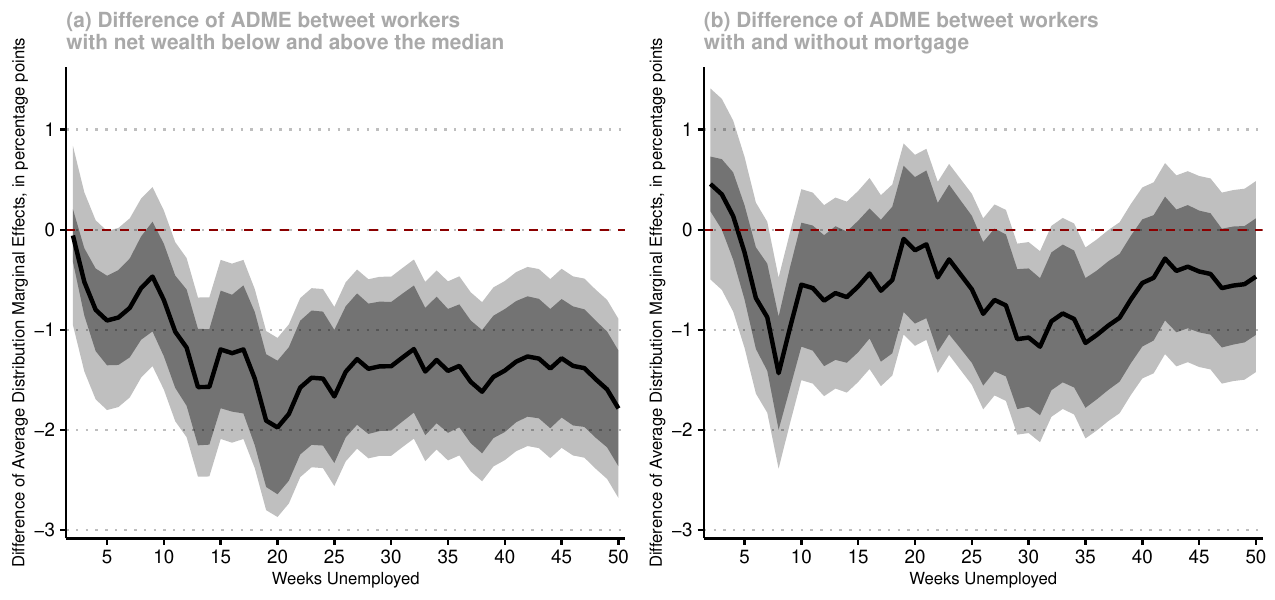}
\end{adjustbox}
\end{center}
\par
\vspace*{-5mm}

\caption{{Difference of estimated average effect of a 10\% increase of unemployment
benefits on the distribution of unemployment duration among liquidity constrained and unconstrained workers . The solid line represents the estimates
while the dark and light shaded areas are the 90\% bootstrapped based
pointwise and simultaneous confidence interval based on 100,000 bootstrap
draws, respectively. Sample size for Panel (a) and (b) is 4,307.}}

\label{fig.amde.het.liquidity}
\end{figure}

In other to further highlight that the effect of UI benefits on unemployment duration is different depending on whether a worker is likely to be liquidity constrained or
not, in Figure \ref{fig.amde.het.liquidity} we plot the difference of the estimated ADME between liquidity constrained and not-liquidity constrained workers, together with their pointwise and simultaneous confidence bands. Panel (a) suggests that UI benefits have a more negative effect on workers with net wealth below the median than on wealthier workers, and that this difference is statistically significant at the 10\% level. Panel (b) reveals a qualitatively similar pattern when one compared workers with mortgage with those without mortgage, though the magnitude of this difference is less pronounced (but we can still reject the null hypothesis that the effects are the same, at the 10\% significance level.)

Taken together, our results provide suggestive evidence that UI benefits have a
non-monotone effect on the unemployment duration distribution and that such
an effect varies whether workers are likely to be liquidity constrained or
not. More precisely, our results suggest that the effect of UI on
unemployment duration is larger for liquidity constrained workers. Through
the lens of the results of \cite{Chetty2008}, our findings
suggest that an increase in UI benefits affects unemployment duration not
only through moral hazard but also because of a \textquotedblleft liquidity
effect.\textquotedblright

\section*{Acknowledgements}
We are most grateful to the editor and two referees for their constructive comments which have lead to a improved paper.
Research funded by Ministerio Econom\'{\i}a y Competitividad (Spain),
ECO2-17-86675-P, and by Comunidad de Madrid
(Spain), MadEco-CM S2015/HUM-3444. Andr\'{e}s~Garc\'{\i}a-Suaza was supported by the Colombia Científica-Alianza EFI Research Program, with code 60185 and contract number FP44842-220-2018, funded by The World Bank through the call Scientific Ecosystems, managed by the Colombian Ministry of Science, Technology and Innovation. Part of this article was written when Pedro H. C. Sant'Anna
was visiting the Cowles Foundation at Yale University, whose hospitality is gratefully acknowledged. 

\bibliography{JMP}

\end{document}